# Atomistically-informed modeling of point defect clustering and evolution in irradiated ThO₂

Sanjoy Kumar Mazumder [a, *], Maniesha Kaur Salaken Singh [b], Tomohisa Kumagai [c], and Anter El-Azab [a]

[a] School of Materials Engineering, Purdue University, West Lafayette, IN 47906, USA

[b] School of Nuclear Engineering, Purdue University, West Lafayette, IN 67906, USA

[c] Central Research Institute of Electric Power Industry (CRIEPI), Yokosuka, Kanagawa, 2400196, Japan.

## Abstract

A cluster dynamics (CD) model has been developed to investigate the nucleation and growth of point defect clusters, i.e., interstitial prismatic loops and nanoscale and sub-nanoscale voids, in ThO₂ during irradiation by energetic particles. The model considers cluster off-stoichiometry due to the asymmetry of point defect generation on the O and Th sublattices under irradiation, as well as the point defect diffusivities and the defect binding energies to clusters. The energies were established using detailed molecular dynamics simulations considering the statistical variability of cluster configuration. A high-order adaptive time-integration has been used to solve the model. The predicted loop density and their average size is in good agreement with reported experimental observations for proton irradiated ThO₂ at 600ºC. The model did not predict void evolution due to the sluggish kinetics of cation vacancies, explaining the absence of voids in proton irradiated ThO₂ (and other oxides) at relatively low temperatures.

* Corresponding author. mazumder@purdue.edu, sanjoymazumder13@gmail.com (Sanjoy Kumar Mazumder)

## 1. Introduction

Thorium dioxide (ThO$_2$) being an actinide oxide has potential application as fuel in nuclear power generation. It has favorable thermophysical properties like lower coefficient of thermal expansion, higher melting point and higher thermal conductivity than uranium dioxide (UO$_2$), which is currently the most common nuclear fuel worldwide [1]. Having adequate thermal conductivity is essential for a nuclear fuel from a heat recovery standpoint [1,2]. However, it is observed that the radiation induced point defects like vacancies and self-interstitial atoms (SIAs) and clusters of these defects, including dislocation loops and voids, impact the thermal transport properties of oxide fuels. Dennett et al. [3] observed a dramatic reduction in thermal conductivity of proton irradiated ThO$_2$ with dose, both at room temperature and 600$^{\circ}$C. Similar reduction in conductivity has also been reported for proton irradiated UO$_2$ and cerium dioxide (CeO$_2$) [4,5]. The scattering of phonons due to nanoscale crystal defects reduce the thermal conductivity in oxides. Thus, it is necessary to accurately measure the concentration of irradiation induced point defects and clusters of all sizes to study their effect on the transport of phonons. Using transmission electron microscopy (TEM) it is possible to obtain the distribution of cluster concentration in the size space. However, point defects and clusters that are smaller than the resolution of TEM, while being most important for understanding thermal transport, pose a significant challenge to TEM observations [3]. An accurate prediction of point defect clustering and evolution with the irradiation time or dose requires a systematic investigation using alternate methods.

Agglomeration or clustering of point defects generated in the displacement cascades during irradiation is studied using cluster dynamics (CD), an approach based on chemical reaction rate theory [6,7] that provides a unified formulation for the nucleation and growth of defect clusters. It uses mean field approximation (MFA) to define rate equations for the temporal evolution of clusters from one size ($n$) to the adjacent ($n \pm 1$) by absorbing or emitting individual point defects. The MFA reduces the spatial distribution of individual defects to average densities at each size thus ignoring any spatial correlation between the point defects and defect clusters. This approach thus assumes that the above interactions take place continuously in space and time. Most CD models consider defects to be generated in the form of Frenkel pairs under irradiation, which drive the evolution of defect clusters. The rate at which clusters of a specific size, $n$, accumulate in the material is obtained as a sum of their production and depletion rates due to the absorption and emission of point defects by similar or different-sized clusters. The rate coefficients of cluster-point defect interactions, i.e., the absorption of SIA (i) and vacancies (v) by dislocation loops ($ni$) and voids ($nv$) given by $\beta_{ni}^i, \beta_{ni}^v, \beta_{nv}^i, \beta_{nv}^v$, respectively, and the emission of SIA and vacancies given by $\alpha_{ni}^i$ and $\alpha_{nv}^v$, respectively, govern the kinetics of the clustering phenomenon. The spectrum of interactions mentioned above are limited by the diffusion of point defects to the immobile defect clusters and microstructural sinks in the material. Consequently, the absorption coefficient, $\beta_{n\theta}^{\theta'}$ of point defects $\theta'$ by clusters $n\theta$ is defined as

$$\beta_{n\theta}^{\theta'} = 2\pi r_{n\theta} Z_{n\theta}^{\theta'} D^{\theta'}, \tag{1}$$

where $D^{\theta'}$ is the diffusivity of point defects $\theta'$, $r_{n\theta}$ is the radius of cluster $n\theta$ and $Z_{n\theta}^{\theta'}$ defines the bias of the cluster for specific point defects. While clusters $n\theta$ can absorb both SIA and vacancies, they can only emit like point defects, i.e., SIA and vacancy clusters can only emit SIAs and vacancies, respectively. The emission coefficient $\alpha_{n\theta}^{\theta}$ is obtained from a detailed mass balance of point defect species between clusters, and is defined as

$$\alpha_{n\theta}^{\theta} = 2\pi r_{(n-1)\theta} \ Z_{(n-1)\theta}^{\theta} D^{\theta} \exp\left(-\frac{E_{n\theta}^b}{k_B T}\right), \tag{2}$$

where $E_{n\theta}^{b}$ is the binding energy of point defect $\theta$ with cluster $(n-1)\theta$, often calculated using molecular dynamics (MD). Thus, the CD model is a system of coupled ODEs governing the evolution of point defects and clusters that are numerically solved using time integrators such as LSODE [8] or CVODE [9], solvers for initial value problems in ODE systems. The general approach of CD discussed above has been used extensively to study defect evolution in materials like Fe [7,10], stainless steel [11,12], Zr [13].

In multicomponent systems, the CD framework has been used to study precipitation kinetics in binary [14,15] and ternary alloys [16]. Another potential application of the CD approach is to predict the defect evolution in irradiated oxide fuels like $CeO_2$, $UO_2$ and $ThO_2$. This has remained relatively less explored. Point defects in oxides such $CeO_2$, $UO_2$ and $ThO_2$ can be charged or neutral. The anion defects can have charges of $\pm 2$, $\pm 1$, or 0, with the plus sign being for vacancies and minus sign for the interstitials, and the cation defects can have charges of $\pm 4$, $\pm 3$, $\pm 2$, $\pm 1$ or 0, with the plus and minus signs for interstitials and vacancies, respectively [17]. The agglomeration of point defects can thus generate clusters of varying stoichiometry and net charge. Considering all possible point defect species and the entire spectrum of clusters would thus complicate CD modeling. Skorek et al. [18] studied the evolution of fission gas bubble in $UO_2$, assuming neutral Schottky defects, i.e., $(U_i, 2O_i)$ and $(V_U, 2V_O)$ defect complexes are generated in the cascade rather than isolated SIA and vacancies of U and O types. They further assumed that Schottky defect complexes are mobile and hence migrates and agglomerates into stoichiometric clusters preserving charge neutrality of the matrix. The clusters trap fission gas atoms, which are otherwise highly mobile, and impede their rate of escape from the surface. In a similar approach, Jonnet et al. [19] modelled the evolution of stoichiometric loops $(U,Pu)O_2$ matrix due to the accumulation of mobile Schottky trio having one $U_i$ and two $O_i$. This straightforward assumption of the clustering of stoichiometric defect complexes fails to address certain fundamental questions later posed by Khalil et al. [20]. Firstly, the asymmetric (or non-stoichiometric) generation of point defects of different species in the displacement cascade due to their varying bonding energies and masses and, secondly, the possibility of point defect agglomeration into clusters having non-stoichiometric compositions owing to a difference in their migration energies. Khalil et al. [20] have presented a more detailed CD model accounting for the evolution of individual monomers, i.e., $O_i$, $U_i$, $V_O$ and $V_U$ and clusters over a wide range of size and stoichiometry. They defined the so-called cluster composition space (CCS) for loops, by counting the respective point defects, i.e., $O_i$, $U_i$ for SIA loops and $V_O$, $V_U$ for vacancy clusters in the fluorite structure of $UO_2$. The modified CD model has the coupled system of evolution equations for monomers and every possible cluster having $m$ U and $n$ O defects in the CCS. This gives a detailed insight into the wide range of stoichiometry of defect clusters and their evolution with time. However, a significant source of error in this model lies in the coefficients, $\alpha_V^{V_U}(m,n)$, $\alpha_V^{V_O}(m,n)$, $\alpha_L^{U_i}(m,n)$, and $\alpha_L^{O_i}(m,n)$ for the emission of $V_U$ and $V_O$ from $(m,n)$ voids (V) and the emission of $U_i$ and $O_i$ from $(m,n)$ SIA loops (L); Khalil et al. used expressions for the binding energies $E_{V_U}^b(m,n)$, $E_{V_O}^b(m,n)$, $E_{U_i}^b(m,n)$, and $E_{O_i}^b(m,n)$ of point defects to the respective clusters that were originally fit for single component metals from MD [21].

In the current work, we investigate the nucleation and growth of dislocation loops and voids in irradiated $ThO_2$ using an atomistically-informed cluster dynamics model. As discussed, a detailed analysis of the non-stoichiometric defect cluster evolution requires a detailed assessment of the binding energy landscape over the CCS. Specifically, the binding energy of defect monomers to clusters having varying stoichiometry have been computed using MD simulations and used as input for the CD model. Point defects $O_i$, $Th_i$, $V_O$ and $V_{Th}$ were assumed to be the only species generated by irradiation and the only mobile species. The diffusivities of these defect monomers, which play an important role in the kinetics of defect clustering, were computed using temperature accelerated MD and self-diffusion data. The CD model was solved using LSODE [8] Package.

## 2. The Cluster Dynamics approach

### 2.1. The cluster composition space of loops and voids

Irradiation of $ThO_2$ generate point defect as Frenkel pairs in the displacement cascade. While point defect monomers $O_i$, $Th_i$, $V_O$ and $V_{Th}$ are introduced via irradiation, they form clusters of a wide range of size and stoichiometry. The cluster composition space (CCS) shows the composition of all possible defect clusters in the fluorite crystal structure of $ThO_2$. Thus, it is important to briefly discuss the crystallography of dislocation loops and voids before building the composition space for the same. SIA loops have been reported on {111} planes of $ThO_2$ having Burgers vector $a/3<111>$ [3,5,22]. Fig. 1(a) and (b) shows the schematic of a (111) SIA loop in $ThO_2$ crystal. It has a layer of Th SIA stacked with two parallel layers of O SIA one above and the other below, along the Th SIA plane normal. Both the O SIA planes have different planar configurations. Like a regular fluorite lattice, the planar distance between Th and O planes in the loop is $a/4\sqrt{3}$, where $a$ is the lattice parameter of $ThO_2$ unit cell. The Th SIAs occupy the octahedral interstitial sites in the Th FCC sublattice as indicated in Fig. 1(d). The sites have a planar co-ordination of close packed {111} planes in FCC. The position of $O_i$ with respect to the $Th_i$ in a SIA loop is similar to that in a perfect crystal. In a perfect $ThO_2$ flourite structure, O occupies the tetrahedral interstitial sites in a Th FCC sublattice as shown in Fig. 1(c) and (d). Thus, each O is surrounded by four Th in a tetrahedral configuration, of which three Th atoms are on a {111} plane and the fourth is on a parallel plane. Similarly, every Th atom is surrounded by eight O atoms in a cubic configuration. While building the CCS for loops, we have gradually populated the octahedral sites with $Th_i$. In order to count the minimum number of $O_i$ required for the loop, every three $Th_i$ arranged in a triangle should have an $O_i$ from one of the two planes together in a close-packed pyramidal configuration. The schematic of an SIA loop having 19 $Th_i$ with the minimum number of $O_i$ is shown in Fig. 2(a). Also, allowing every $Th_i$ in the loop to have eight surrounding O atoms give the maximum number of $O_i$ that the loop can accommodate, as shown in Fig. 2(b). While building the CCS for loops, $O_i$ are progressively added to the cluster having minimum number of $O_i$ until the coordination number of all the $Th_i$ in the loop is fulfilled.

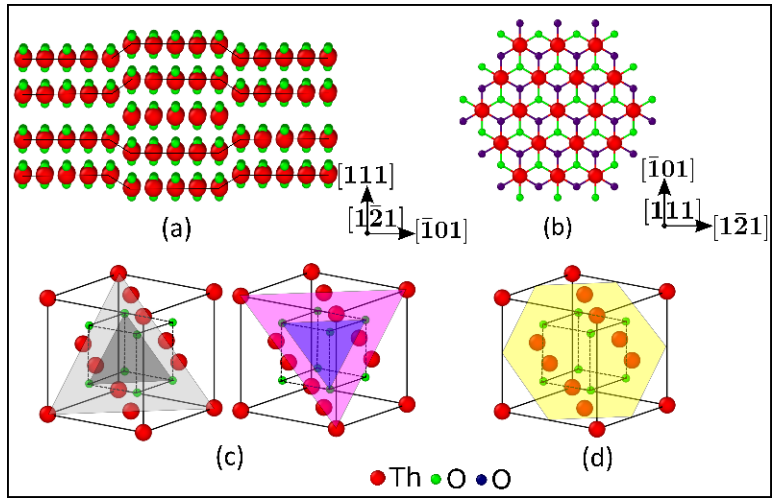

Fig. 1. Schematic of (111) SIA loop in ThO$_2$. (a) Stacking of the loop between (111) planes as viewed along ($1\bar{2}1$) axis. (b) The SIA loop structure shows a (111) planar FCC configuration of Th SIAs bonded to O SIAs, marked with two different colors indicating their location above or below the Th SIA plane. (c) The Th FCC sublattice and O cubic sublattice in ThO$_2$ fluorite unit cell is indicated by solid and dotted lines, respectively. (d) The center and the vertices of the shaded plane indicates the octahedral interstitial sites in FCC Th sublattice.

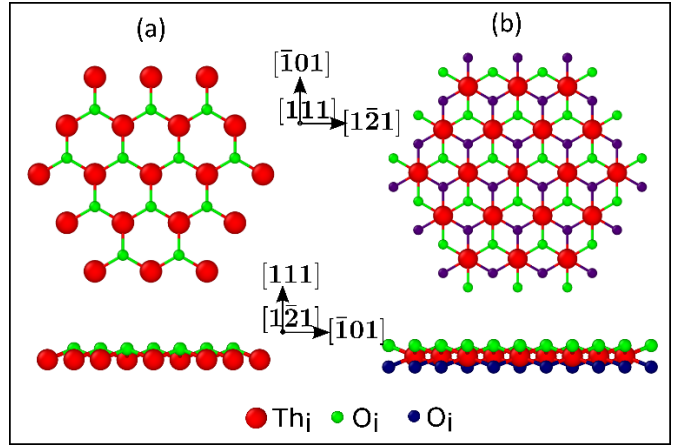

Fig. 2. An SIA loop having 19 Th SIA with (a) minimum and (b) maximum number of Th-O bonds.

Fig. 3(a) shows the CCS for SIA loops having up to 40 $Th_i$. Every point in the CCS indicates a loop having a particular composition of $Th_i$ ($n$) and $O_i$ ($m$). The loops having stoichiometric composition, i.e., $m/n = 2$, is indicated by the black solid line in fig 3(a). Thus, the possible range of composition appears both in the hypo and hyper-stoichiometric regimes of the CCS. It is to be noted that the loops represented in the CCS has a hexagonal geometry in the {111} plane of the $ThO_2$ lattice. The voids are 3D defects which are built by removing Th and O atoms from the lattice progressively. In order to build a void having minimum number of $V_O$ corresponding to a specific number of $V_{Th}$, we have removed one O atom for every four Th atoms arranged in a tetrahedral configuration. Similarly, removing a Th atom with its eight surrounding O atoms give a void having maximum number of $V_O$. Fig 4(a) and (b) shows the geometry of a void having 13 $V_{Th}$ with a minimum and maximum number of $V_O$ respectively. The CCS for the voids as shown in Fig. 3(b) is obtained by progressively removing O atoms from the void having minimum number of $V_O$. The voids having stoichiometric composition is indicated by the black solid line in the CCS. Similar to the loops, the range of void composition appears both in the hypo and hyper-stoichiometric regimes of the CCS. Any specific cluster in the CCS for both loops and voids is indicated as ($n, m$) where $n$ and $m$ indicates the number of Th and O monomers respectively.

Knowing the composition space, we can write the equations for evolution of loops and voids for all possible cluster composition.

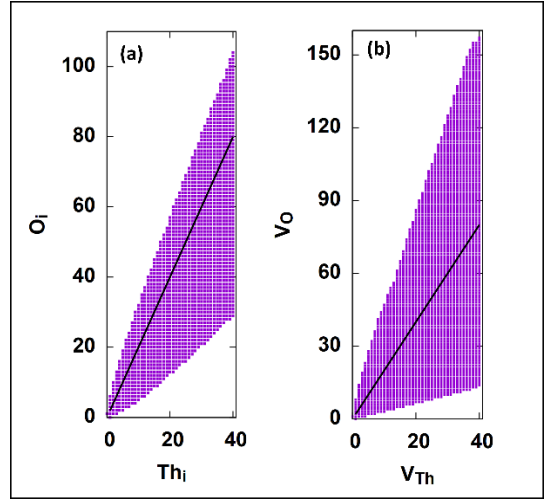

Fig. 3. Cluster composition space (CCS) for (a) SIA loops and (b) voids. The black solid line indicates clusters having stoichiometric composition.

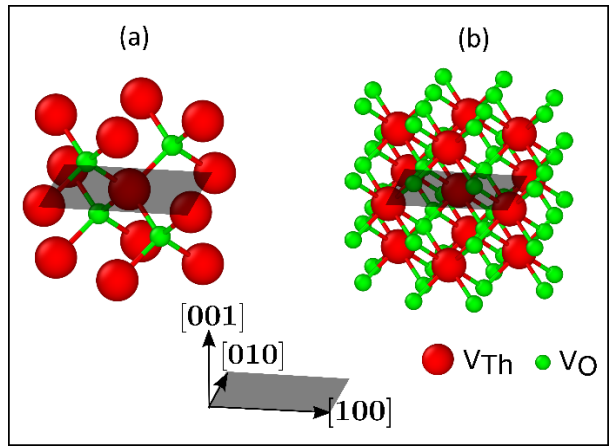

Fig. 4. A void having 13 $V_{Th}$ with (a) minimum and (b) maximum number of $V_O$.

## 2.2. The cluster dynamics model

The modified CD model has a coupled system of ODEs denoting the evolution of individual monomers, SIA loops and voids of all possible sizes compositions. Any cluster in the CCS den8526oted by $(\boldsymbol{n}, \boldsymbol{m})$ can either grow into $(\boldsymbol{n+1}, \boldsymbol{m})$ and $(\boldsymbol{n}, \boldsymbol{m+1})$ clusters or shrink into $(\boldsymbol{n-}$ $\mathbf{1}, \boldsymbol{m}$) and ($\boldsymbol{n}, \boldsymbol{m} - \mathbf{1}$) clusters by absorbing or emitting point defects of Th and O. At the boundary, reactions resulting in the generation of clusters outside the CCS should be neglected. The density of loops and voids having composition ($\boldsymbol{n}, \boldsymbol{m}$) is denoted by $\boldsymbol{C_L(n, m)}$ and $\boldsymbol{C_V(n, m)}$, respectively. The temporal evolution of these densities is given by the following equations [20]:

$$
\begin{aligned}
\frac{dC_L(n,m)}{dt} = {} & \beta_L^{Th_i}(n-1,m)C_{Th_i}C_L(n-1,m) \\
& + \beta_L^{O_i}(n,m-1)C_{O_i}C_L(n,m-1) \\
& + \left[\beta_L^{V_{Th}}(n+1,m)C_{V_{Th}} + \alpha_L^{Th_i}(n+1,m)\right]C_L(n+1,m) \\
& + \left[\beta_L^{V_O}(n,m+1)C_{V_O} + \alpha_L^{O_i}(n,m+1)\right]C_L(n,m+1) \\
& - \left[\alpha_L^{Th_i}(n,m) + \beta_L^{V_{Th}}(n,m)C_{V_{Th}} + \beta_L^{Th_i}(n,m)C_{Th_i}\right]C_L(n,m) \\
& - \left[\alpha_L^{O_i}(n,m) + \beta_L^{V_O}(n,m)C_{V_O} + \beta_L^{O_i}(n,m)C_{O_i}\right]C_L(n,m)
\end{aligned}
\tag{3}
$$

$$
\begin{aligned}
\frac{dC_V(n,m)}{dt} = {} & \beta_V^{V_{Th}}(n-1,m)C_{V_{Th}}C_V(n-1,m) \\
& + \beta_V^{V_O}(n,m-1)C_{V_O}C_V(n,m-1) \\
& + \left[\beta_V^{Th_i}(n+1,m)C_{Th_i} + \alpha_V^{V_{Th}}(n+1,m)\right]C_V(n+1,m) \\
& + \left[\beta_V^{O_i}(n,m+1)C_{O_i} + \alpha_V^{V_O}(n,m+1)\right]C_V(n,m+1) \\
& - \left[\alpha_V^{V_{Th}}(n,m) + \beta_V^{Th_i}(n,m)C_{Th_i} + \beta_V^{V_{Th}}(n,m)C_{V_{Th}}\right]C_V(n,m) \\
& - \left[\alpha_V^{V_O}(n,m) + \beta_V^{O_i}(n,m)C_{O_i} + \beta_V^{V_O}(n,m)C_{V_O}\right]C_V(n,m),
\end{aligned}
\tag{4}
$$

where $\boldsymbol{\beta_T^{\theta}(n, m)}$ and $\boldsymbol{\alpha_T^{\theta}(n, m)}$ are the coefficients of absorption and emission respectively of point defects $\boldsymbol{\theta} = Th_i, O_i, V_{Th}, V_O$ by any cluster T, SIA loop (L) or void (V), having a composition ($n, m$) where $\boldsymbol{n + m > 2}$. The ($\boldsymbol{n}, \boldsymbol{m}$) SIA loops can form due to the absorption of $Th_i$ and $O_i$ by $(n-1, m)$ and $(n, m-1)$ loops, respectively. The rate of these reactions is given by the first and the second terms in Eq. (3). The successive terms in the equation represents the rate of generation of $(n, m)$ loops due to the absorption of $V_{Th}$ and $V_O$ and the emission of $Th_i$ and $O_i$ by $(n+1, m)$ and $(n, m+1)$ loops, respectively. Finally, the depletion rate of $(n, m)$ clusters due to similar interactions with point defects is given by the remaining terms in the equation. Eq. (4) gives an equivalent expression for the evolution of voids. It should be mentioned that clusters denoted by (1, 0) and (0, 1) indicates Th and O point defects respectively. Corresponding equations for point defect and dimer evolution [20], given by Eq. (5) - (10) below, are solved together with Eq. (3) and (4) in the CD model.

The time rate of change of the point defect densities, $C_{Th_i}$, $C_{V_{Th}}$, $C_{O_i}$, and $C_{V_O}$ and the dimers $C_L(1,1)$ and $C_V(1,1)$ are given by:

$$
\begin{aligned}
\frac{dC_{Th_i}}{dt} = {} & G_{Th_i} - R_{iv}^{Th} C_{Th_i} C_{V_{Th}} - \beta_L^{O_i}(1,0) C_{O_i} C_{Th_i} + \beta_L^{V_O}(1,1) C_{V_O} C_L(1,1) \\
& - C_{Th_i} \sum_{n \geq 1, m \geq 1} \left( \beta_L^{Th_i}(n,m) C_L(n,m) + \beta_V^{Th_i}(n,m) C_V(n,m) \right) \\
& + \sum_{n \geq 1, m \geq 1} \alpha_L^{Th_i}(n,m) C_L(n,m)
\end{aligned}
\tag{5}
$$

$$
\begin{aligned}
\frac{dC_{V_{Th}}}{dt} = {} & G_{V_{Th}} - R_{iv}^{Th} C_{Th_i} C_{V_{Th}} - \beta_V^{V_O}(1,0) C_{V_{Th}} C_{V_O} + \beta_V^{O_i}(1,1) C_{O_i} C_V(1,1) \\
& - C_{V_{Th}} \sum_{n \geq 1, m \geq 1} \left( \beta_V^{V_{Th}}(n,m) C_V(n,m) + \beta_L^{V_{Th}}(n,m) C_L(n,m) \right) \\
& + \sum_{n \geq 1, m \geq 1} \alpha_V^{V_{Th}}(n,m) C_V(n,m)
\end{aligned}
\tag{6}
$$

$$\frac{dC_{O_i}}{dt} = G_{O_i} - R_{iv}^O C_{O_i} C_{V_O} - \beta_L^{Th_i}(0,1)C_{O_i}C_{Th_i} + \beta_L^{V\,Th}(1,1)C_{V_{Th}}C_L(1,1) \tag{7}$$

$$- C_{O_i} \sum_{n \geq 1, m \geq 1} \left( \beta_L^{O_i}(n,m)C_L(n,m) + \beta_V^{O_i}(n,m)C_V(n,m) \right)$$

$$+ \sum_{n \geq 1, m \geq 1} \alpha_L^{O_i}(n,m)C_L(n,m)$$

$$\frac{dC_{V_O}}{dt} = G_{V_O} - R_{iv}^O C_{O_i} C_{V_O} - \beta_V^{V\,Th}(0,1)C_{V_{Th}}C_{V_O} + \beta_V^{Th_i}(1,1)C_{Th_i}C_V(1,1) \tag{8}$$

$$- C_{V_O} \sum_{n \geq 1, m \geq 1} \left( \beta_V^{V_O}(n,m)C_V(n,m) + \beta_L^{V_O}(n,m)C_L(n,m) \right)$$

$$+ \sum_{n \geq 1, m \geq 1} \alpha_V^{V_O}(n,m)C_V(n,m)$$

$$\frac{dC_L(1,1)}{dt} = \beta_L^{Th_i}(0,1)C_{Th_i}C_{O_i} + \left[ \beta_L^{V\,Th}(2,1)C_{V_{Th}} + \alpha_L^{Th_i}(2,1) \right]C_L(2,1) \tag{9}$$

$$+ \left[ \beta_L^{V_O}(1,2)C_{V_O} + \alpha_L^{O_i}(1,2) \right]C_L(1,2)$$

$$- \left[ \alpha_L^{Th_i}(1,1) + \beta_L^{V\,Th}(1,1)C_{V_{Th}} + \beta_L^{Th_i}(1,1)C_{Th_i} \right]C_L(1,1)$$

$$- \left[ \alpha_L^{O_i}(1,1) + \beta_L^{V_O}(1,1)C_{V_O} + \beta_L^{O_i}(1,1)C_{O_i} \right]C_L(1,1)$$

$$\frac{dC_V(1,1)}{dt} = \beta_V^{V\,Th}(0,1)C_{V_{Th}}C_{V_O} + \left[ \beta_V^{Th_i}(2,1)C_{Th_i} + \alpha_V^{V\,Th}(2,1) \right]C_V(2,1) \tag{10}$$

$$+ \left[ \beta_V^{O_i}(1,2)C_{O_i} + \alpha_V^{V_O}(1,2) \right]C_V(1,2)$$

$$- \left[ \alpha_V^{V\,Th}(1,1) + \beta_V^{Th_i}(1,1)C_{Th_i} + \beta_V^{V\,Th}(1,1)C_{V_{Th}} \right]C_V(1,1)$$

$$- \left[ \alpha_V^{V_O}(1,1) + \beta_V^{O_i}(1,1)C_{O_i} + \beta_V^{V_O}(1,1)C_{V_O} \right]C_V(1,1).$$

The terms $G_{Th_i}, G_{V_{Th}}, G_{O_i}, G_{V_O}$ in Eq. (5) – (8) are the rate of point defect generation in the displacement cascade. The rate of recombination of Th and O SIA and vacancies is denoted by $R_{iv}^O$ and $R_{iv}^{Th}$, respectively. The depletion of point defect density results from absorption by clusters, recombination and dimer generation, i.e., loops and voids having composition (1,1). Whereas they are generated due to dissociation of dimers, emission by clusters and most importantly in the displacement cascade at a constant damage rate. Thus, the CD model can be defined for any number of clusters from the CCS and solved together with the rate equations for point defect evolution.

### 2.3. Parameters of the cluster dynamics model

As mentioned previously, the different reaction rate coefficients govern the evolution kinetics of the defects. The SIA-vacancy recombination is limited by migration of individual defects towards each other. Thus, $R_{iv}^S$ where $S$ = Th, O is defined as:

$$R_{iv}^S = 4\pi r_{iv}^S \big(D_{S_i} + D_{V_S}\big), \tag{11}$$

where $r_{iv}^S \sim 2.5a$ is the radius of a spherical region around any SIA or vacancy within which the recombination reaction would occur spontaneously and $a$ is the lattice parameter of ThO$_2$. $D_{S_i}$ and $D_{V_S}$ denote the diffusivity of point defects at the temperature of irradiation. Similarly, the coefficient of point defect absorption by voids and loops also depend on the diffusivities and the cluster radii [20],

$$\beta_V^{V_S}(n,m) = \frac{4\pi R_V(n,m)D_{V_S}}{1 + \frac{a}{R_V(n,m)}} \tag{12}$$

$$\beta_V^{S_i}(n,m) = \frac{4\pi R_V(n,m)D_{S_i}}{1 + \frac{a}{R_V(n,m)}}. \tag{13}$$

In Eq. (12) and (13), $\beta_V^{V_S}(n,m)$ and $\beta_V^{S_i}(n,m)$ gives the coefficient of point defect absorption by voids having composition $(n,m)$. $R_V$ is the void radius and is defined for a spherical void as

$$R_{\mathrm{V}}(n,m) = \left(\frac{3(n+m)\Omega}{4\pi}\right)^{1/3}, \tag{14}$$

where $\Omega$ denotes the atomic volume and is simply considered equal to $a^3/12$. In addition, the coefficient of absorption by SIA loops, $\beta_{\mathrm{L}}^{\mathrm{V_s}}(n,m)$ and $\beta_{\mathrm{L}}^{\mathrm{S_i}}(n,m)$ also depend on its bias towards individual defects given by the factors $Z_{\mathrm{L}}^{\mathrm{V_s}}(n,m)$ and $Z_{\mathrm{L}}^{\mathrm{S_i}}(n,m)$ [20],

$$\beta_{\mathrm{L}}^{\mathrm{V_s}}(n,m) = \left(\frac{4\pi R_{\mathrm{L}}(n,m)Z_{\mathrm{L}}^{\mathrm{V_s}}(n,m)D_{\mathrm{V_s}}}{1+\frac{a}{R_{\mathrm{L}}(n,m)}}\right) \tag{15}$$

$$\beta_{\mathrm{L}}^{\mathrm{S_i}}(n,m) = \left(\frac{4\pi R_{\mathrm{L}}(n,m)Z_{\mathrm{L}}^{\mathrm{S_i}}(n,m)D_{\mathrm{S_i}}}{1+\frac{a}{R_{\mathrm{L}}(n,m)}}\right). \tag{16}$$

The radius of SIA loops is defined assuming them to be circular. Although our CCS is built for hexagonal loops, an assumption of the circular shape simplifies the model.

$$R_{\mathrm{L}}(n,m) = 0.3215 \cdot a \cdot \sqrt{n}. \tag{17}$$

The expression of loop radius is obtained by equating the area of a hexagonal loop with a circle of radius $R_{\mathrm{L}}(n,m)$. From Eq. (17), we see that the radius varies as the square root of only the number of $\mathrm{Th_i}$, $n$, in the loop. This is because, the $\mathrm{Th_i}$ being considerably larger than $\mathrm{O_i}$ determines the loop size. The Burgers vector $b$ of SIA loops on $\{111\}$ plane is equal to $a/\sqrt{3}$. The bias factors are defined like Duparc et al. [7] and is shown below,

$$Z_{\mathrm{L}}^{\mathrm{V_s}}(n,m) = Z_{\mathrm{V_s}}^{\mathrm{d}} + \frac{\left[\left(\frac{b}{8\pi a}\right)^{\frac{1}{2}} z_{\mathrm{V_s}} - Z_{\mathrm{V_s}}^{\mathrm{d}}\right]}{(n+m)^{\gamma_{\mathrm{V_s}}/2}} \tag{11}$$

$$Z_L^{S_i}(n,m) = Z_{S_i}^d + \frac{\left[\left(\frac{b}{8\pi a}\right)^{\frac{1}{2}} z_{S_i} - Z_{S_i}^d\right]}{(n+m)^{\frac{r_{S_i}}{2}}}. \tag{19}$$

The terms $\boldsymbol{Z_{V_S}^d}$ and $\boldsymbol{Z_{S_i}^d}$ denote the bias factors due to straight dislocations and is equal to 1 for $V_{Th}$ and $V_O$ and 1.2 for $Th_i$ and $O_i$. The remaining factors of Eq. (5) and (6) are given as $\boldsymbol{z_{V_S} = 35}$ for both $V_{Th}$ and $V_O$ and $\boldsymbol{z_{S_i} = 42}$ for $Th_i$ and $O_i$. Finally, as mentioned in Eq. (2), the emission of point defect from clusters is obtained from a detailed mass balance, equating the flux of like point defect species out of a specific $(\boldsymbol{n, m})$ cluster and into the next cluster $(\boldsymbol{n + 1, m})$ or $(\boldsymbol{n, m + 1})$ in the CCS. We have only considered the emission of like point defects from clusters, i.e., $Th_i$ and $O_i$ from loops and $V_{Th}$ and $V_O$ from voids. In addition to the mobility, emission coefficient is also governed by the change in energy associated with the binding of respective point defects to clusters [20],

$$\alpha_L^{S_i}(n,m) = \beta_L^{S_i}(n,m) \cdot \exp\left(-\frac{E_{S_i}^b(n,m)}{k_B T}\right) \tag{20}$$

$$\alpha_L^{V_S}(n,m) = \beta_L^{V_S}(n,m) \cdot \exp\left(-\frac{E_{V_S}^b(n,m)}{k_B T}\right). \tag{21}$$

In Eq. (20) and (21), $\boldsymbol{E_{V_{Th}}^b(n,m)}$ and $\boldsymbol{E_{V_O}^b(n,m)}$ are the binding energies of $V_{Th}$ and $V_O$ to voids whereas $\boldsymbol{E_{Th_i}^b(n,m)}$ and $\boldsymbol{E_{O_i}^b(n,m)}$ are the binding energy of $Th_i$ and $O_i$ to SIA loops, respectively. Thus, a detailed definition of the reaction rate coefficients helps us identify the crucial parameters of the modified CD model, i.e., the point defect diffusivities and the binding energies of individual point defects to different clusters. In the following subsections, calculation of these parameters has been discussed in details.

### 2.3.1. Calculation of cluster binding energies

In a single component system, the change in energy associated with the binding of a point defect to a cluster is given as [7,21]

$$E_{n\theta}^{\mathrm{b}} = E_{\theta}^{\mathrm{f}} - \left(E_{n\theta}^{\mathrm{f}} - E_{(n-1)\theta}^{\mathrm{f}}\right), \tag{22}$$

where $E_{\theta}^{\mathrm{f}}$ and $E_{n\theta}^{\mathrm{f}}$ are the formation energies of point defects and defect clusters respectively. Thus, the binding energy is defined with respect to the different point defect species in $ThO_2$ as:

$$E_{\mathrm{Th_i}}^{\mathrm{b}}(n,m) = E_{\mathrm{Th_i}}^{\mathrm{f}} - \left(E_{\mathrm{L}}^{\mathrm{f}}(n,m) - E_{\mathrm{L}}^{\mathrm{f}}(n-1,m)\right) \tag{23}$$

$$E_{\mathrm{O_i}}^{\mathrm{b}}(n,m) = E_{\mathrm{O_i}}^{\mathrm{f}} - \left(E_{\mathrm{L}}^{\mathrm{f}}(n,m) - E_{\mathrm{L}}^{\mathrm{f}}(n,m-1)\right) \tag{24}$$

$$E_{\mathrm{V_{Th}}}^{\mathrm{b}}(n,m) = E_{\mathrm{V_{Th}}}^{\mathrm{f}} - \left(E_{\mathrm{V}}^{\mathrm{f}}(n,m) - E_{\mathrm{V}}^{\mathrm{f}}(n-1,m)\right) \tag{25}$$

$$E_{\mathrm{V_O}}^{\mathrm{b}}(n,m) = E_{\mathrm{V_O}}^{\mathrm{f}} - \left(E_{\mathrm{V}}^{\mathrm{f}}(n,m) - E_{\mathrm{V}}^{\mathrm{f}}(n,m-1)\right). \tag{26}$$

The formation energy of clusters having composition $(\boldsymbol{n}, \boldsymbol{m})$, i.e., $E_{\mathrm{L}}^{\mathrm{f}}(\boldsymbol{n}, \boldsymbol{m})$ and $E_{\mathrm{V}}^{\mathrm{f}}(\boldsymbol{n}, \boldsymbol{m})$ can be further defined by

$$E_{\mathrm{L}}^{\mathrm{f}}(n,m) = E_{\mathrm{i}}^{\mathrm{Random}}(n,m) + E_{\mathrm{L}}^{\mathrm{Cluster}}(n,m) \tag{27}$$

$$E_{\mathrm{V}}^{\mathrm{f}}(n,m) = E_{\mathrm{v}}^{\mathrm{Random}}(n,m) + E_{\mathrm{V}}^{\mathrm{Cluster}}(n,m). \tag{28}$$

where $E_{\mathrm{i}}^{\mathrm{Random}}(n,m)$ denote the energy required to create a random distribution of $n$ $Th_i$ and $m$ $O_i$ in a perfect $ThO_2$ lattice. The distributed SIAs are sufficiently far apart to not interact with each other. $E_{\mathrm{L}}^{\mathrm{Cluster}}(n,m)$ is the clustering energy of the randomly distributed SIAs into a loop. Thus, the definition of $E_{\mathrm{L}}^{\mathrm{f}}(n,m)$ and $E_{\mathrm{L}}^{\mathrm{Cluster}}(n,m)$ of a loop differs in the reference configuration, i.e., the perfect lattice for the former and a crystal with a random distribution of SIAs for the latter. Similarly, $E_{\mathrm{v}}^{\mathrm{Random}}(n,m)$ denotes the energy required to create a random distribution of $n$ $V_{Th}$ and $m$ $V_O$ and $E_{\mathrm{V}}^{\mathrm{Cluster}}(n,m)$ is the clustering energy of vacancies into a void. We can further define $E_{\mathrm{i}}^{\mathrm{Random}}(n,m)$ and $E_{\mathrm{v}}^{\mathrm{Random}}(n,m)$ as:

$$E_i^{Random}(n, m) = n \cdot E_{Th_i}^f + m \cdot E_{O_i}^f \tag{29}$$

$$E_v^{Random}(n, m) = n \cdot E_{V_{Th}}^f + m \cdot E_{V_O}^f, \tag{30}$$

where $E_{Th_i}^f, E_{O_i}^f, E_{V_{Th}}^f$ and $E_{V_O}^f$ are the formation energy of respective point defects. Substituting Eq. (29) and (30) into Eq. (27) and (28), respectively and further substituting $E_L^f(n, m)$ and $E_V^f(n, m)$ into Eq. (23) – (26), we obtain

$$E_{Th_i}^b(n, m) = E_L^{Cluster}(n - 1, m) - E_L^{Cluster}(n, m) \tag{31}$$

$$E_{O_i}^b(n, m) = E_L^{Cluster}(n, m - 1) - E_L^{Cluster}(n, m) \tag{32}$$

$$E_{V_{Th}}^b(n, m) = E_V^{Cluster}(n - 1, m) - E_V^{Cluster}(n, m) \tag{33}$$

$$E_{V_O}^b(n, m) = E_V^{Cluster}(n, m - 1) - E_V^{Cluster}(n, m). \tag{34}$$

Thus, the binding energy of an SIA or a vacancy to a loop or void $(n, m)$, respectively, is obtained as the difference in the clustering energy of consecutive loops or voids in the CCS. $E_L^{Cluster}(n, m)$ and $E_V^{Cluster}(n, m)$ are defined as the energy change associated with the clustering of a random distribution of respective point defects. Using LAMMPS[23], we compute the clustering energies $E_L^{Cluster}(n, m)$ and $E_V^{Cluster}(n, m)$ for loops and voids respectively. The SIA loops discussed in section 2.1 is introduced in a perfect $ThO_2$ simulation box, having a dimension of $16 \times 16 \times 16$ fluorite unit cell with 16384 Th atoms and 32768 O atoms, between two {111} planes as shown in Fig. 1(a). The simulation box with the loop is relaxed by conjugate gradient minimization using an EAM type interatomic potential for actinide oxides developed by Cooper et al. [24] Following relaxation, the system is equilibrated in an NPT ensemble at 10 K and 0 bar pressure for 60 ps. This allows the system to undergo thermal relaxation at a low temperature. A high temperature equilibration might destabilize the loop and dissociate it depending on the loop stoichiometry. A final relaxation is carried out on the equilibrated system and the total internal energy, $E^{Loop}$, is obtained as an output. The initial and final configuration of the (111) planes near the SIA loop are shown in Fig. 5. The equilibrated structure in Fig. 5(b) shows significant distortion of the (111) planes due to the presence of the SIA loop. In another simulation box of $\mathbf{16 \times 16 \times 16}$ $ThO_2$ unit cells, an equal number of $Th_i$ and $O_i$ as the SIA loop is randomly placed in the octahedral interstitial positions of the $ThO_2$ lattice, at least 5-unit cells apart. It was similarly relaxed using conjugate gradient minimization, equilibrated at 10 K and 0 bar in an NPT ensemble and finally relaxed again. The total internal energy, $\mathbf{E^{Random}}$, is obtained as an output. The clustering energy $\mathbf{E_L^{Cluster}(n, m)}$ is calculated from the total internal energy of the two systems as:

$$E_L^{Cluster}(n, m) = E^{Loop} - E^{Random}. \tag{35}$$

A positive value of $\mathbf{E_L^{Cluster}(n, m)}$ indicates that the clustering is not favourable energetically. A plot of $\mathbf{E_L^{Cluster}(n, m)}$ of all the loops in the CCS is shown in Fig. 6(a). We see that $\mathbf{E_L^{Cluster}(n, m)}$ is positive and close to zero for small and hypo-stoichiometric loops and becomes more negative with increased size and hyper-stoichiometry.

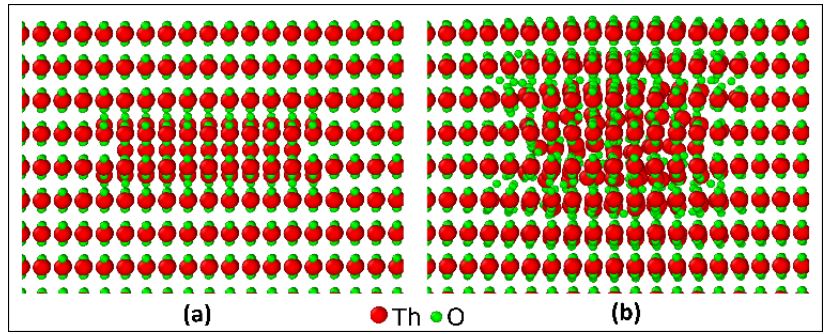

Fig. 5. (a) Initial configuration of the planes near a (111) SIA loop having 19 $Th_i$ and 54 $O_i$. (b) Significant distortion of (111) planes near the SIA loop after NPT equilibration at 10K and 0 bar.

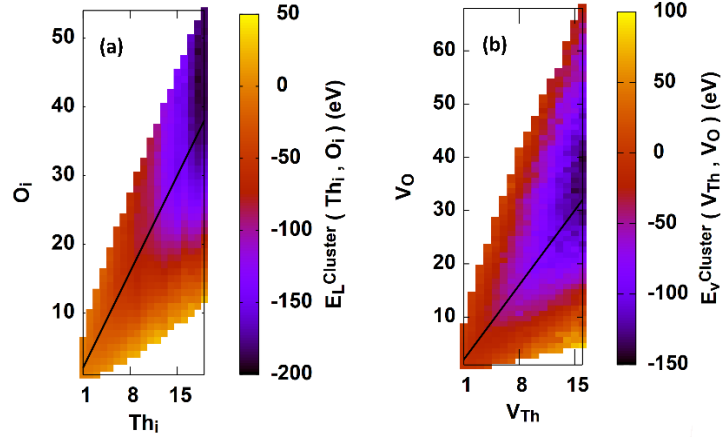

Fig. 6. The clustering energy of (a) SIA loops and (b) voids shown in their respective CCS. The black solid line indicates the stoichiometric composition of clusters.

The clustering energy of spherical voids, $E_V^{\text{Cluster}}(n, m)$, is computed in MD similar to the SIA loops. The voids described in section 2.1 is introduced in a perfect simulation cell of dimension $16 \times 16 \times 16$ ThO$_2$ unit cells and relaxed with conjugate gradient minimization. This is followed by thermal equilibration in an NPT ensemble at 10 K and 0 bar pressure. After a final relaxation, the total internal energy of the system with the void, $E^{\text{Void}}$ is obtained as output. In another simulation box, randomly distributed V$_{\text{Th}}$ and V$_{\text{O}}$ are equilibrated similarly and the total internal energy, $E^{\text{Random}}$ is obtained. Fig. 6(b) shows the $E_V^{\text{Cluster}}(n, m)$ of different voids in the CCS, calculated using Eq. (33). The clustering energy calculated for both the SIA loops and voids show small fluctuations in some region of the CCS. These fluctuations can be averaged to obtain a smooth variation of the mean energy values.

The loops considered in section 2.1 are hexagonal, determined by the arrangement of Th$_i$ in the octahedral interstitial sites as shown in Fig. 1(d). While fulfilling the coordination number of all the Th$_i$ in the hexagonal loop, we can arrange the O$_i$ in different possible configurations. The clustering energy of the loops, as shown in Fig. 6(a) have however been calculated considering a single configuration of the O$_i$ in a loop of specific composition $(n, m)$. Also, the small loops can reasonably exist in different geometry in the cascade due to the arrangement of $Th_i$ in different configurations. They are not strictly restricted to assume a geometry corresponding to the minimum clustering energy, $E_L^{Cluster}(n, m)$. Thus, it is important to explore the configuration space of small loops and calculate their $E_L^{Cluster}(n, m)$. The small loops are built on the plane of octahedral interstitial sites shown in Fig. 1(d). Starting with a $Th_i$ in an octahedral site, the first nearest neighboring sites are identified and populated one $Th_i$ at a time to obtain a set of $Th_i$ dimers. Again, the first nearest neighbors of each dimer are identified to place the third $Th_i$ and form a trimer. Thus, we obtain a set of $Th_i$ trimers corresponding to a single $Th_i$ dimer and proceed similarly to build all possible configurations of a $Th_i$ 4-mer. Identifying the neighbors of a specific $Th_i$ $n$-mer and building all possible $(n + 1)$-mers generate several configurations that are exactly similar and can be obtained from each other by simple symmetry operations, i.e., translation, rotation and inversion about specific axes. There exists a definite number of distinct isomers of each $Th_i$ $n$-mer that increases with the size $n$ combinatorially. The $O_i$ are added to the $Th_i$ $n$-mers in a way similar to that in the hexagonal loops. The minimum number of $O_i$ for an $n$-mer is obtained by placing them in sites such that each $Th_i$ has an $O_i$ at a distance equal to a Th - O bond length in a perfect fluorite structure. $O_i$ are progressively added to this SIA cluster until the coordination number of all the $Th_i$ is fulfilled and we obtain an $n$-mer having the maximum stoichiometry.

It is reasonable to expect that large clusters would adopt a compact configuration, i.e., hexagonal or circular due to capillary effects. Since, the number of isomers increase combinatorially with the size of $n$-mer, it is expensive to explore all possible configuration of small clusters beyond $Th_i$ 6-mers. The clustering energy, $E_L^{Cluster}(n, m)$ of the clusters thus generated is calculated using MD as mentioned above and is shown in Fig. 7. Each configuration of a $Th_i$ $n$-mer is represented by a single line, i.e., blue, red, orange and green for $n = 3, 4, 5$ and $6$ respectively, varying from the minimum to maximum possible stoichiometry. The trend of $E_L^{Cluster}(n, m)$ is similar to that of SIA loops as shown in Fig. 6(a), i.e., there exists a minimum near the stoichiometric cluster composition. Also, we obtain a range of $E_L^{Cluster}(n, m)$ for any $(n, m)$ SIA cluster based on its configuration as shown in Fig. 7(b). This range is narrow for hypo-stoichiometric clusters and gradually widens with an increase in cluster stoichiometry.

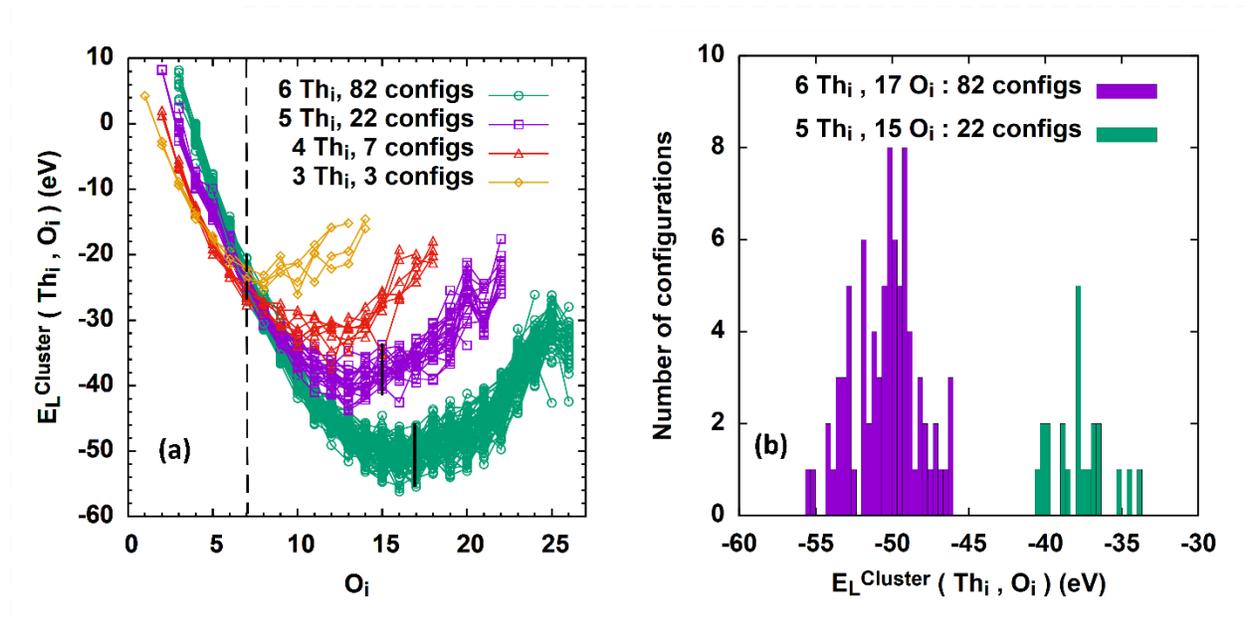

Fig. 7. (a) Clustering energy of $(n, m)$ SIA clusters having $n$ = 3, 4, 5 and 6 $Th_i$ and varying stoichiometry. (b) Histogram plot of $E_L^{Cluster}$ over all configurations in (5 $Th_i$, 15 $O_i$) and (6 $Th_i$, 17 $O_i$) loops. The black solid line in (a) marks the spread of $E_L^{Cluster}$ over the possible configurations in the loops mentioned.

It is also interesting to note that small clusters have low $E_L^{Cluster}(n, m)$ compared to the large ones in the hypo-stoichiometric regime, whereas it is opposite for the hyper-stoichiometric clusters. The dashed line in Fig. 7(a) indicates the point where this transition in the trend of $E_L^{Cluster}(n, m)$ with cluster size take place. From the data in Fig. 7(a), average value of $E_L^{Cluster}(n, m)$ corresponding to $n$ = 3, 4, 5 and 6 is obtained over the entire range of stoichiometry. Finally, the $E_L^{Cluster}(n, m)$ for large loops from $n$ = 7 to 19 plotted in Fig. 6(a) and the average values for small loops are fitted with separate parabolic expressions, varying with $m$, for every $n$. Fig. 8(a) shows the fitted $E_L^{Cluster}(n, m)$ plots with the calculated data. The red solid line marks the clusters having minimum $E_L^{Cluster}(n, m)$ whereas the black line indicates the ones having stoichiometric composition, for every $n$. $E_L^{Cluster}(n, m)$ is defined according to the parabolic expression shown below:

$$E_L^{Cluster}(n, m) = x_n \cdot (m - m_{min})^2 + E_{min} \qquad (36)$$

where $E_{min}$ indicates the minimum in the $E_L^{Cluster}(n, m) - m$ plot for a particular $n$ and $m_{min}$ is the value of $m$ corresponding to $E_{min}$. In other words, for clusters having $n$ Th$_i$, $m_{min}$ is the number of O$_i$ present in the cluster with the minimum $E_L^{Cluster}$. Thus, $E_{min}$, $m_{min}$ and the coefficient of the parabolic term in Eq. (34), $x_n$ are all functions of $n$ that has been plotted and fitted in Fig. 9(a), (b) and (c) respectively. The energy $E_{min}$ varies with $n^{1.3}$ whereas the parameter $x_n$, which indicates the convexity of the parabolic term in Eq. (34) at a fixed $n$, decreases monotonically as $n^{-0.25}$ and approaches zero for large values of $n$. Substituting these equations fitted for the unknown terms back into Eq. (34), we obtain the expression for $E_L^{Cluster}(n, m)$.

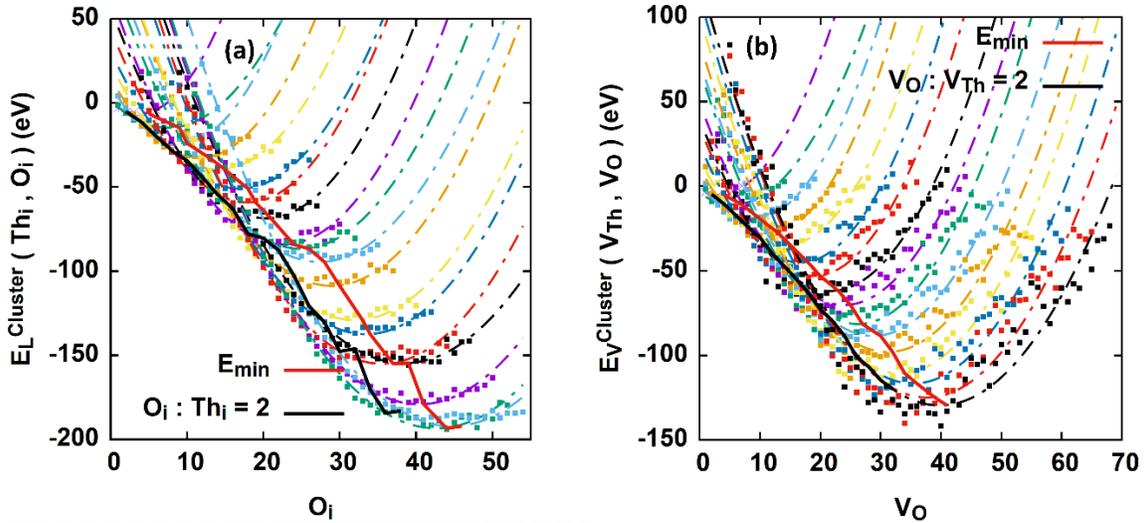

Fig. 8. Plot of the average (a) loop and (b) void clustering energies with the number of O$_i$ and V$_O$ respectively for every $n$.

We can further calculate $E_{Th_i}^b(n, m)$ and $E_{O_i}^b(n, m)$ using Eq. (31) – (34). A plot of $E_{O_i}^b(n, m)$ and $E_{Th_i}^b(n, m)$ on the CCS of loops is shown in Fig. 10(a) and (b) respectively. A negative value of $E_{O_i}^b(n, m)$ implies that it is energetically favourable for the loop $(n, m)$ to form a $(n, m-1)$ loop by emitting an $O_i$. In other words, more negative the value of $E_{O_i}^b(n, m)$ is, larger is the emission coefficient $\alpha_{O_i}(n, m)$ in Eq. (20). In Fig. 10(a), we can clearly see that $E_{O_i}^b(n, m)$ is negative for hyper-stoichiometric clusters. Similarly, a negative value of $E_{Th_i}^b(n, m)$ implies that the emission of $Th_i$ by a $(n, m)$ loop to form a $(n-1, m)$ loop is favourable energetically. Fig. 10(b) shows evidently that the loops with hypo-stoichiometric composition have a negative $E_{Th_i}^b(n, m)$.

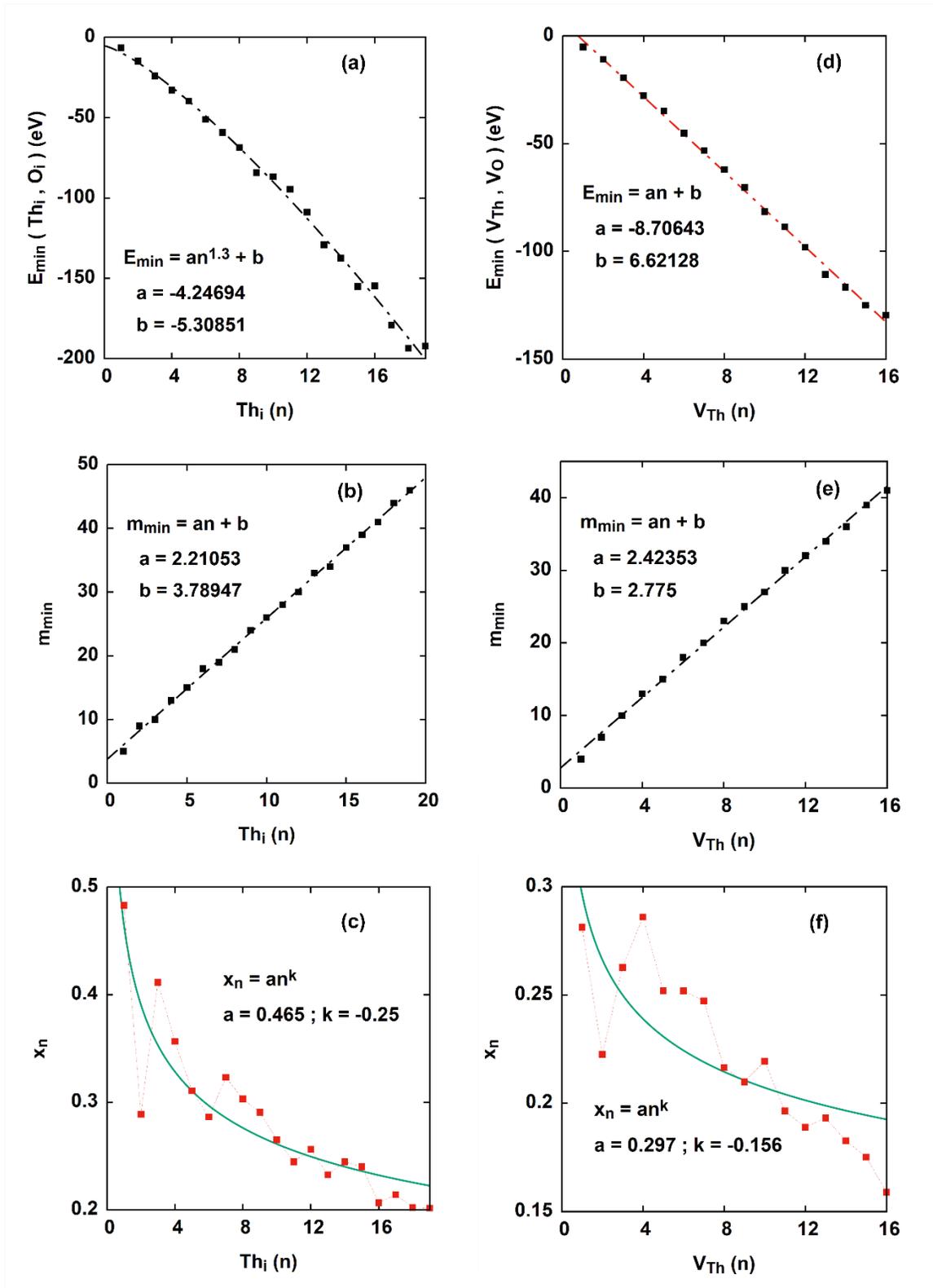

Fig. 9. Fitted equations for the plots of $E_{min}$, $m_{min}$ and $x_n$ with $n$, in (a), (b), (c) loops and (d), (e), (f) voids.

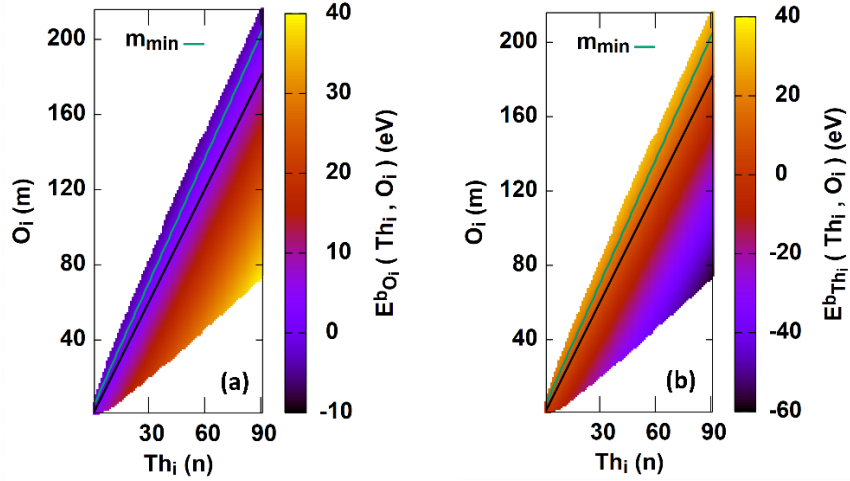

Fig. 10. The binding energy of (a) $O_i$ and (b) $Th_i$ to loops shown in the CCS. The black solid line indicates the stoichiometric composition of loop.

The clustering energy of spherical voids, $E_V^{Cluster}(n, m)$, is computed in MD similar to the SIA loops. The voids described in section 2.1 is introduced in a perfect simulation cell of dimension $16 \times 16 \times 16$ $ThO_2$ unit cells and relaxed with conjugate gradient minimization. This is followed by thermal equilibration in an NPT ensemble at 10 K and 0 bar pressure. After a final relaxation, the total internal energy of the system with the void, $E^{Void}$ is obtained as output. In another simulation box, randomly distributed $V_{Th}$ and $V_O$ are equilibrated similarly and the total internal energy, $E^{Random}$ is obtained. Fig. 6(b) shows the $E_V^{Cluster}(n, m)$ of different voids in the CCS, calculated using Eq. (33). The fluctuations in the energy values can be averaged to obtain a smooth energy landscape over the CCS. Similar to SIA clusters, small voids can also assume different geometry in the cascade. However, exploring the configuration space of voids in a manner similar to the SIA clusters is more expensive owing to their 3D geometry. Instead, we have assumed the voids to be spherical and fitted an expression similar to Eq. (34) for $E_V^{Cluster}(n, m)$ in the CCS. The fitting of $E_{min}, m_{min}$ and $x_n$ for the voids is shown in Fig. 9 (d), (e) and (f) respectively. Unlike the loops, $E_{min}$ decreases linearly with $n$ whereas $x_n$ approaches zero asymptotically varying with $n^{-0.156}$. Also, the binding energy of $V_O$ and $V_{Th}$ to a void $(n, m)$ in the CCS, calculated using Eq. (34) and (33) shows that the emission of $V_O$ and $V_{Th}$ by hyper and hypo stoichiometric voids respectively are energetically more favorable due to the negative values of $E^b_{V_O}(n, m)$ and $E^b_{V_{Th}}(n, m)$ as shown in Fig. 11(a) and (b). The expressions for clustering energy of loops and voids are used as input parameter of the CD model.

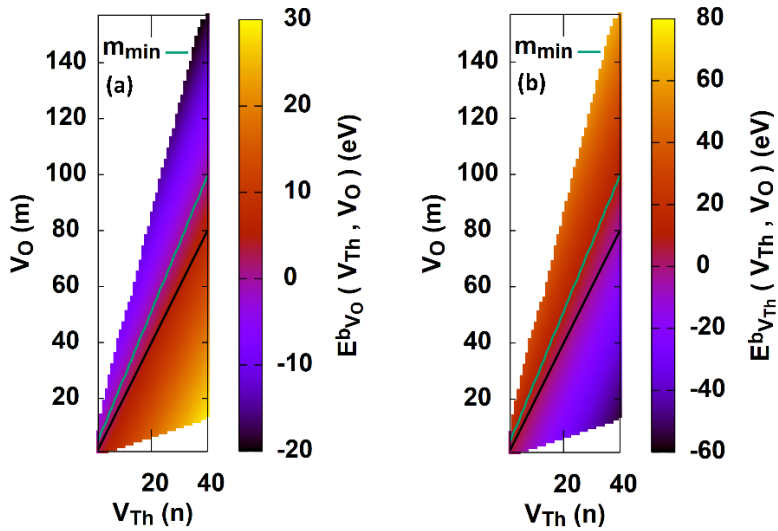

Fig. 11. The binding energy of (a) $V_O$ and (b) $V_{Th}$ to voids shown in the CCS. The black solid line indicates the stoichiometric composition of voids.

### 2.3.2. Calculation of point defect diffusivities

The diffusivity of point defects $Th_i$, $O_i$, $V_{Th}$ and $V_O$ indicated by $\boldsymbol{D_{Th_i}}, \boldsymbol{D_{O_i}}, \boldsymbol{D_{V_{Th}}}$ and $\boldsymbol{D_{V_O}}$ respectively is another important parameter of the CD model. Accelerated MD techniques like temperature accelerated MD or TAMD [25], hyperdynamics [26,27] and parallel replica dynamics are widely used to study the diffusion of atomic species and defects. We have used TAMD in which an $O_i$ was introduced in an octahedral interstitial site of the perfect $ThO_2$ lattice having dimension $10 \times 10 \times 10$ $ThO_2$ flourite unit cells. With the EAM potential by Cooper et al. [24], the simulation cell is relaxed using conjugate gradient minimization. The system with the point defect is then equilibrated in an NVT ensemble at an elevated temperature, 2700 K, for 1 ns. Using the 'Wigner Seitz defect Analysis' modifier in OVITO visualization tool [28], we detect the position of $O_i$ in the cell after every ps. The atom ID of $O_i$ detected at every ps is different which indicates exchange of the $O_i$ with regular lattice O atom, an interstitialcy mechanism of diffusion where a cation atom in an octahedral interstitial site kicks out a cation atom from a regular lattice site which in turn moves to the nearest octahedral site. Xiao et al. [29] used climbing image nudged elastic band (CI-NEB) method in DFT to compute the migration energy of $O_i^{-2}$ and reported the interstitialcy mechanism to be more favourable. Knowing the position at every ps, we compute the mean squared displacement, MSD of $O_i$ and plot it against time. The slope of MSD plot gives a measure of the diffusivity using the Einstein's relation:

$$< r^2 > = 6Dt, \qquad (37)$$

with $\boldsymbol{D}$ being the diffusivity of the point defect under consideration. The TAMD run for $O_i$ was repeated 10 times at 9 different temperatures. The $\boldsymbol{D_{O_i}}$ from different runs and the mean $\boldsymbol{D_{O_i}}$ at each temperature are plotted against inverse temperature. The slope and the intercept of the linearly fitted mean $\boldsymbol{D_{O_i}}$ gives the migration energy $\boldsymbol{E_{O_i}^m}$ and pre-exponential diffusivity factor $\boldsymbol{D_{O_i}^0}$ respectively. We have obtained a value of 1.577 eV for $\boldsymbol{E_{O_i}^m}$ and 0.266 cm$^2$/s for $\boldsymbol{D_{O_i}^0}$. The migration energy barrier of $O_i^{-2}$ reported by Xiao et al. [29] is 1.04 eV. Thus, the interatomic potential by Cooper et al. predicts a higher energy barrier for $O_i$ migration. Similarly, we have studied the diffusivity of $V_O$, by removing an O atom from its regular lattice site and used TAMD as discussed above. The $\boldsymbol{E_{V_O}^m}$ obtained is 0.564 eV which is almost equal to the value reported by Aidhy [30], i.e., 0.563 V, using the nudged elastic band method as implemented in LAMMPS with the EAM potential developed by Cooper et al. On the other hand, CI-NEB calculations in DFT shows that $E_{V_O}^m$ is highly anisotropic with large values of about 1.27 eV, Yun et al. [31], 2.16 eV, Xiao et al. [32] and 1.97 eV, Liu et al. [33]

The diffusivity of $Th_i$ in $ThO_2$ has not yet been studied unlike cation interstitial diffusion in other actinide oxides namely $U_i$ in $UO_2$ [34] and $Ce_i$ in $CeO_2$ [35]. The predominant mechanism considered for interstitial migration in both the oxides are direct diffusion from one octahedral site to the adjacent and an interstitialcy mechanism. NEB calculations in DFT yield an energy barrier of $7.9 - 8.8$ eV and $4.1 - 4.7$ eV for the two mechanisms of $U_i$ migration in $UO_2$ respectively [34]. For $Ce_i$ in $CeO_2$, the migration energy barrier is 6.31 eV and 4.19 eV respectively [35]. TAMD simulations performed for $Th_i$ in $ThO_2$ and subsequent analysis using OVITO indicates an interstitialcy mechanism as the atom ID detected for $Th_i$ every ps are different. The $E_{Th_i}^m$ obtained is 3.172 eV which indicates that TAMD with the EAM potential by Cooper et al. yields a smaller migration barrier for $Th_i$ diffusion than that predicted for cation diffusion in other oxides by NEB using DFT.

The energy barrier associated with migration of $V_{Th}$ from one lattice site to the nearest has been reported to be 4.47 eV by Yun et al. [31] and 6.67 eV by Xiao et al. [36] using climbing image nudged elastic band (CI-NEB) method in DFT. However, TAMD simulations with the EAM potential by Cooper et al. [24] could not capture the diffusion of $V_{Th}$ even at 2800 K. This implies that the kinetics of Th vacancies in $ThO_2$ is highly sluggish. In order to have a reasonable estimate, we have calculated $D_{V_{Th}}$ using the following expression for self-diffusivity of **Th** atoms in $ThO_2$:

$$D_{Th}^{self}(T) = C_{Th_i}(T) \cdot D_{Th_i}(T) + C_{V_{Th}}(T) \cdot D_{V_{Th}}(T)$$

$$\Rightarrow D_{Th}^{self}(T) = C_{Th_i}(T) \cdot D_{Th_i} + C_{V_{Th}}(T) \cdot D_{V_{Th}}^o \cdot \exp\left(-\frac{E_{V_{Th}}^m}{k_B T}\right) \tag{38}$$

where $D_{Th}^{self}$ indicates the self-diffusivity of Th atoms in ThO$_2$. Experimental studies on Th diffusion in ThO$_2$ by King [37], report the values of $D_{self}^{o}$ and $E_{self}^{m}$ to be 0.35 cm$^2$/s and 6.495 eV respectively. The high activation energy of Th self-diffusion in ThO$_2$ was later confirmed by H. Matzke [38]. $C_{Th_i}(T)$ and $C_{V_{Th}}(T)$ are the equilibrium concentration of **Th** SIA and vacancies at any temperature $T$. Thermodynamic calculations performed by Maniesha et al. [39] gives the equilibrium concentration of defect species in ThO$_2$. Thus, knowing $D_{Th}^{self}$, $D_{Th_i}$, $C_{Th_i}$ and $C_{V_{Th}}$ at two different temperatures we can compute $D_{V_{Th}}^{o}$ and $E_{V_{Th}}^{m}$, which are the pre-exponential diffusivity factor and the migration energy barrier for V$_{Th}$ diffusion in ThO$_2$. The $D_{Th_i}$ obtained using TAMD has been used in Eq. (36) for the calculation of $D_{V_{Th}}$. The values of $D_{V_{Th}}^{o}$ and $E_{V_{Th}}^{m}$ thus obtained are 85.64 cm$^2$/s and 4.837 eV respectively. The migration energy of V$_{Th}$ thus obtained is very close to 4.47 eV, the $E_{V_{Th}}^{m}$ reported by Yun et al. [31] *Figure S1.* in supplementary information shows the plot of monomer diffusivities obtained from TAMD.

## 3. Cluster Dynamics Model Results

The binding energy of point defects to clusters and the diffusivity of point defects computed above have been used in the CD model to study the evolution of clusters in irradiated ThO$_2$. Recently, Dennett, Deskins and Khafizov et al. [40] has reported observing a distribution of faulted interstitial Frank loops on {111} planes in the BF TEM images of ThO$_2$ irradiated with protons at 600ºC and upto 0.79 dpa. They measured the density of loops over varying sizes and the average loop radius at different irradiation doses. Also, voids and vacancy clusters were not observed even at the highest irradiation dose reported. The purpose of our CD model is to reproduce this experimentally observed density distribution of loops at similar irradiation conditions. With atomistic simulations, we have obtained a reasonable set of parameters to solve the model.

However, the uncertainties associated with the determined parameters provide a range over which we can vary them to obtain the best fit of the model result with the TEM observations. An important input to the CD model is the generation rate for different point defect species, i.e., $O_i$, $V_O$, $Th_i$ and $V_{Th}$. Experimentally, the $ThO_2$ samples were irradiated with a proton flux of $\mathbf{1.8 \times 10^{13}}$ ions/cm²s. From the reported ion fluence, we have calculated the exposure time corresponding to different irradiation doses. The total displacements of Th and O when divided by the irradiation time gives the total dose rate. The ratio of average plateau dose of Th and O in the depth - damage distribution profile obtained from SRIM is 0.457. Using this ratio and the total dose rate, the generation rate of $O_i$ and $V_O$ Frenkel pairs is obtained as $\mathbf{1.138 \times 10^{-6}}$ dpa/s and that of $Th_i$ and $V_{Th}$ Frenkel pairs is 5.203 ×10⁻⁷ dpa/s.

We have checked the sensitivity of our CD model to the input parameters by varying them within their range of uncertainity to better fit the density distribution of loops obtained from irradiation experiments. An estimate of the uncertainty in $E_{O_i}^b$ and $E_{Th_i}^b$ is calculated by analyzing the $E_L^{Cluster}(n, m)$ of loops from Fig. 9(a) in the near-stoichiometric region of the cluster composition space. The distribution of $E_L^{Cluster}(n, m)$ over the possible range of cluster configurations define the uncertainty, $\Delta E$, associated with it. Fig. 9(a) shows that $\Delta E$ is maximum for the hyper-stoichiometric composition of the loops. The distribution of $E_L^{Cluster}(n, m)$ over all possible configurations for (5 $Th_i$, 15 $O_i$) and (6 $Th_i$, 17 $O_i$) loops are shown in Fig. 9(b). $\Delta E$ is measured from the spread of $E_L^{Cluster}$ in the histogram plots which are ∼7 ev and ∼9 eV for the loops of two different sizes respectively. Also, $\Delta E$ is comparatively small for hypo-stoichiomeric loops which can be clearly seen from Fig. 9(a). As the number of configurations increase for progressively larger loops, $\Delta E$ increases as well. Assuming this to be true for all clusters in the CCS, we can define an effective $\Delta E \sim 10$ eV, about the average $E_L^{\text{Cluster}}(n,m)$ for all loops fitted using Eq. (34).

$$\text{Actual } E_L^{\text{Cluster}}(n,m) = E_L^{\text{Cluster}}(n,m) \pm \frac{\Delta E}{2}. \qquad (39)$$

Consequently, the definition of $E_{\text{Th}_i}^{b}$ and $E_{\text{O}_i}^{b}$ using Eq. (31) and (32) allows us to define the uncertainty associated with them as

$$\text{Actual } E_{\text{Th}_i}^{b}(n,m) = E_{\text{Th}_i}^{b}(n,m) \pm \Delta E \qquad (40)$$

$$\text{Actual } E_{\text{O}_i}^{b}(n,m) = E_{\text{O}_i}^{b}(n,m) \pm \Delta E. \qquad (41)$$

Thus, we can vary $E_{\text{Th}_i}^{b}$ and $E_{\text{O}_i}^{b}$ within their range of uncertainty defined by Eq. (40) and (41) to better reproduce the experimental observations. In the simplest possible approach, we have multiplied $E_{\text{Th}_i}^{b}$ and $E_{\text{O}_i}^{b}$ with a factor of 2 across the CCS.

The kinetics of loop growth is also influenced by the mobility of point defect species in ThO$_2$. In order to analyse the sensitivity of our model to the diffusivity of Th$_i$, O$_i$, V$_{\text{Th}}$ and V$_{\text{O}}$ we have varied the respective migration energies slightly from their reference values. Result from a series of CD runs with varying migration energies has been presented in *Figures S2* to *S4*. in the supplementary information. The optimized set of migration energies that yield the best fit of loop size distribution from CD with the TEM observations is listed in Table 1. The results obtained from CD are least sensitive to the diffusivity of V$_{\text{Th}}$ whereas it is significant for the others.

Table 1 Optimized migration energies of point defects

| $E_{\text{O}_i}^{m}$ | $E_{\text{V}_\text{O}}^{m}$ | $E_{\text{Th}_i}^{m}$ | $E_{\text{V}_\text{Th}}^{m}$ |
|---|---|---|---|
| 1 eV | 1.5 eV | 2.3 eV | 4.837 eV |

Using the calculated point defect generation rate, the updated $E_{Th_i}^b$ and $E_{O_i}^b$ and the optimized point defect diffusivities, the CD model is solved for the evolution of clusters at 600°C. We observed the evolution of loops only in the near-stoichiometric region of the CCS. The reason for this preferential evolution in the composition space is that the binding energy of $Th_i$ and $O_i$, i.e., $E_{Th_i}^b$ and $E_{O_i}^b$ are both positive for the loops in this region, making their emission difficult and hence the loops more stable. This allowed us to truncate the CCS and consider only the loops having near-stoichiometric compositions. Consequently, it improves the computational efficiency of the CD model by reducing the number of equations to be solved. Fig. 12 shows the evolution of SIA loops in the truncated CCS where the black solid line denotes the stoichiometric loop compositions. The distribution of density over varying sizes or the size distribution function (SDF) of loops at different doses obtained from CD are shown in Fig. 13 (a). We see at any dose, there is a peak in loop density for a specific size. With an increase in irradiation dose, this peak shifts to larger sizes accompanied with a reduction in peak density indicating growth of SIA loops. The SDF obtained from CD is compared with the TEM observations by Dennett, Deskins and Khafizov et al. [40] as shown in Fig. 13(b). It is clear that the distribution observed at 0.16 dpa and 0.47 dpa in TEM is closely predicted by CD above a loop radius of 1.5 nm. The discrepancy in the measured densities of smaller loops and that predicted using CD may be attributed to the limited resolution of the TEM used for observations. Also, the measured SDF at 0.79 dpa is noticeably different from that predicted using CD. Though CD predicts a shift in the density peak to larger average loop sizes, TEM measurements show otherwise. This is clear from the plot of average loop radius with dose as shown in Fig. 14 (a). Secondly, the peak loop density observed at 0.79 dpa is higher than the peak density at 0.16 and 0.47 dpa which is in contradiction to the evolution of loop SDF predicted by CD. Similar to CD predictions, evolution of large voids was not observed in TEM even at high irradiation doses predominantly owing to the sluggish mobility of vacancies. However, vacancy monomers, dimers and very small voids have been predicted to accumulate in the matrix.

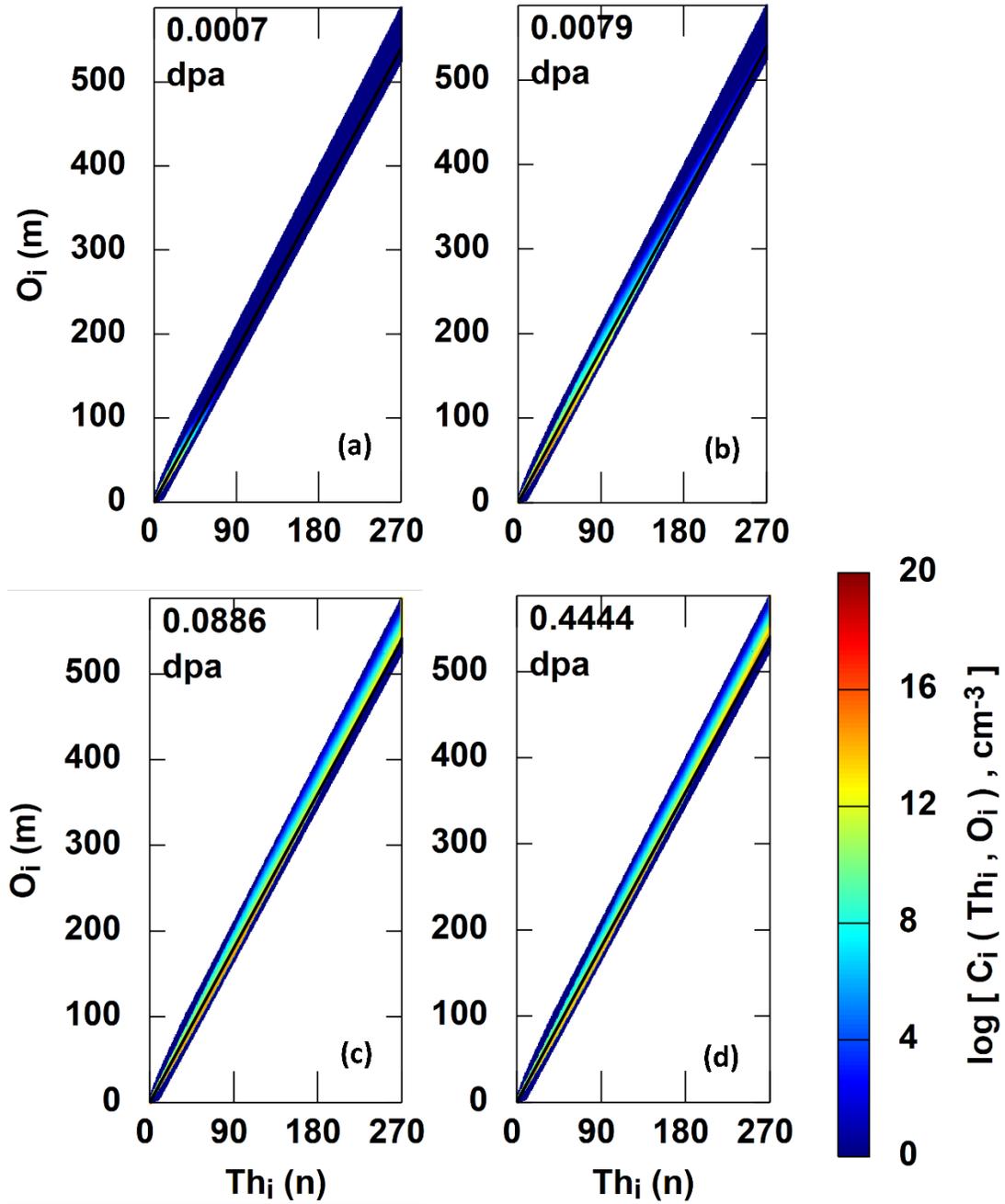

Fig. 12. Evolution of SIA loops in the truncated CCS obtained with $2 \times E_{Th_i}^b$, $2 \times E_{O_i}^b$ and the updated $E_{O_i}^m$, $E_{Th_i}^m$, $E_{V_O}^m$ at (a) 0.0007 (b) 0.0079 (c) 0.0886 and (d) 0.4444 dpa respectively. the solid black line indicates clusters having stoichiometric composition.

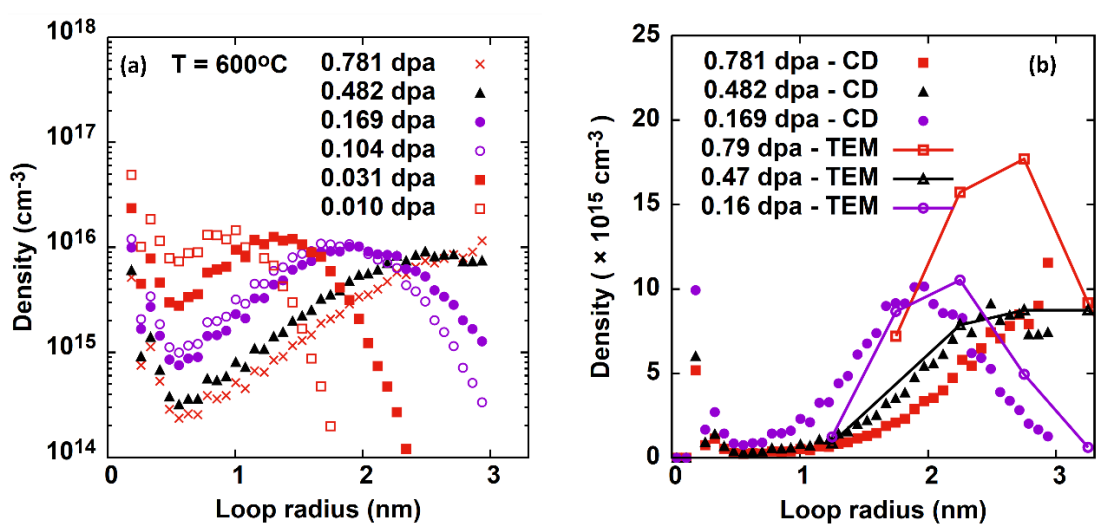

Fig. 13. (a) Size distribution function (SDF) of SIA loops at different doses obtained from CD. (b) SDF of SIA loops obtained from TEM observations compared with CD results.

The CD prediction of average loop radius increases with irradiation dose as shown in Fig. 14(a). It is also evident that the predicted radii lie within the error bound of TEM measurements. The total loop density as shown in Fig. 14(b) increases very slowly to almost reach saturation with dose. The loop density measured from TEM is less than that predicted by CD. This is expected because the resolution of TEM limits its accuracy in measuring the density of small loops. Fig. 14(b) also shows the evolution of loops having varying stoichiometry, $S = m/n$. It is evident that the density of hypo-stoichiometric loops dominates over stoichiometric and hyper-stoichiometric loops at low doses. However, at high irradiation doses, the growth of hyper-stoichiometric loops is higher. The variation of monomer and dimer densities is shown in Fig. 14(c). After an initial accumulation, we see a depletion in $O_i$ density primarily owing to their migration and clustering with $Th_i$ to form dimers and small loops. Annihilation of $O_i$ with $V_O$ also contribute slightly to the reduction in their density. The density of $Th_i$ on the other hand shows a slight depletion while predominantly accumulating with dose. This is expected due to the sluggish kinetics of $Th_i$ as compared to $O_i$ which is the most mobile species at 600°C. The $V_O$ also accumulate in the matrix until they start migrating to form vacancy dimers and small vacancy clusters. The annihilation of $Th_i$ and $V_{Th}$ starts as the $Th_i$ become mobile in the matrix. However, $V_{Th}$ being the most sluggish defect keep accumulating even at high irradiation doses. Fig. 14(d) shows the evolution of vacancy monomer and dimer density with a linear variation of irradiation dose to compare with the experimental observations. Dennett, Deskins and Khafizov et al. [40] attempted to measure the density of charged oxygen vacancies in the system from the changes in the electronic structure of the lattice induced by irradiation. Spectroscopic ellipsometry measure the electronic transitions which is obtained as a complex dielectric function. Amplitude of the imaginary part in the dielectric function increases with irradiation dose as shown in Fig. 14(d), indicating an enhanced density of charged defects. The modified CD model discussed in this work assumes that the irradiated $ThO_2$ matrix only has neutral oxygen vacancies and predicts a depletion in its density with dose, unlike the observed trend of increasing charged defect density. It also predicts a constant increase in the density of $V_{Th}$ and vacancy dimers as shown in Fig. 14(d). Thus, a correlation exists between the spectroscopy data and the density of vacancy dimers predicted by CD which can only be explained if the dimers and very small clusters carry effective electronic charge which contributes to the observed amplitude in the complex dielectric function.

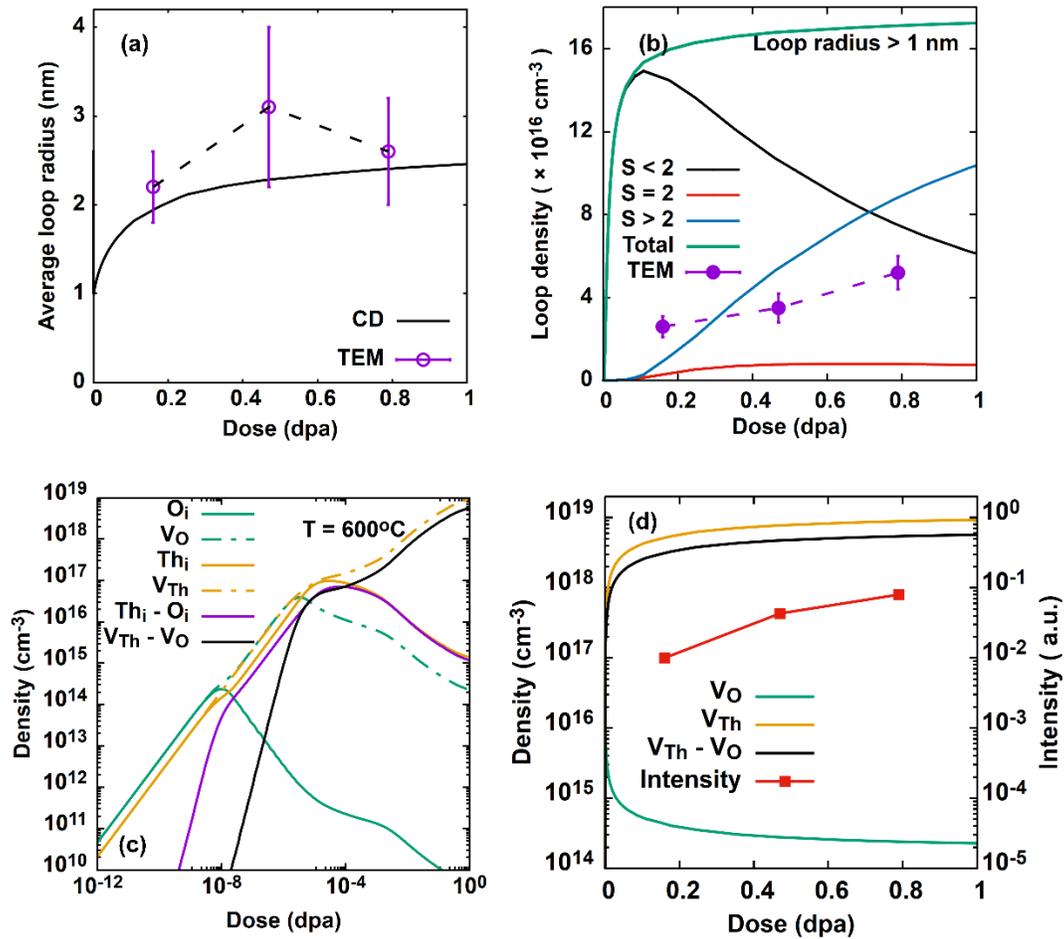

Fig. 14. (a) Comparison of the average loop radius at different doses obtained from CD with TEM measurements. (b) Density evolution of loops, having varying stoichiometry ( $S = m/n$ ), with dose. Total loop density measured with TEM at different doses is also plotted. (c) Evolution of monomer and dimer density. (d) Density of vacancy monomers and dimers plotted with irradiation dose. Amplitude of the imaginary part of complex dielectric function obtained from spectroscopic ellipsometry indicating the density of charged vacancy defects, plotted with irradiation dose.

The effect of irradiation temperature on the evolution of loops was also studied using our modified CD model. Fig 15(a) and (b) shows the SDF evolution of loops at 400°C and 700°C respectively. At low temperature, CD predicts a limited loop growth due to sluggish monomer kinetics thus accumulating small loops and SIA clusters in the matrix. Whereas, at 700°C the SDF reaches even higher loop radii as compared to that at 600°C, shown in Fig. 13(c). We see a comparison of the average loop radius with dose at different temperatures in Fig. 15(c). Also, the peak loop density at different dose reduces with an increase in temperature. This is primarily because the annihilation of monomers due to recombination increases as they become more mobile at high temperature. It is also evident from the plot of total loop density shown in Fig. 15(d). At 400°C, accumulation of small loops continuously increases its total density with dose. As temperature increases and the monomers become mobile, significant loop growth is observed due to which the total loop density almost saturate as the dose increases. The rate of recombination also accelerates at high temperature limiting the total density of loops to even lower values. Also, as observed in Fig. 14(b) the loops are predominantly hypo-stoichiometric, i.e., $S < 2$ at low dose and hyper-stoichiometric, i.e., $S > 2$ at high doses. This transition from hypo to hyper-stoichiometric loop compositions take place at low doses as temperature increase and is clearly shown in Fig. 15(d). However, at low temperatures the loops remain predominantly hypo-stoichiometric even at high irradiation doses. Void evolution was not predicted even at a high temperature of 900°C.

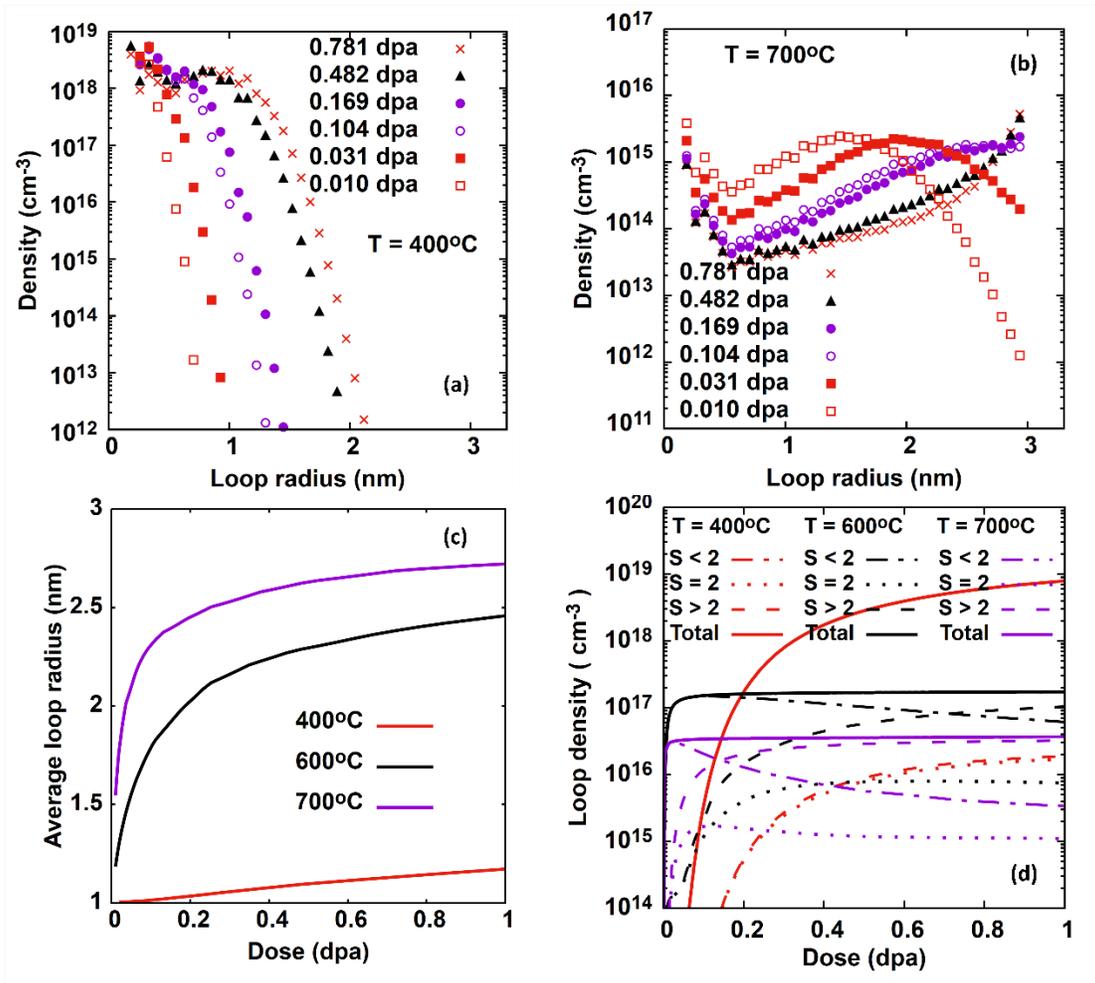

Fig. 15. SDF of loops predicted by CD at (a) 400ºC and (b) 700ºC. (c) The variation of average loop radius with dose at different temperatures. (d) The density variation of hypo-stoichiometric, stoichiometric, and hyper-stoichiometric loops with dose at different temperatures where $S = m/n$. The variation of total loop density with dose at different temperatures is also plotted.

Finally, we also plotted the variation of monomer density in the matrix at different temperatures as shown in Fig. 16(a) and (b). The depletion of $O_i$ density in the matrix begins with the clustering of SIA to form loops. As temperature increases, the depletion begins at a much lower dpa owing to the higher mobility of $O_i$. Similarly, the depletion of other monomers begins at low dose as they become sufficiently mobile to cluster and form loops. Fig. 16(c) plots the mean stoichiometry of SIA loops, $S_{mean}$, obtained as a ratio of the total $O_i$ density to the total $Th_i$ density clustered in loops.

$$S_{\text{mean}} = \left[ \sum_{(m,n)} m \cdot C_{\text{L}}(m,n) \right] / \left[ \sum_{(m,n)} n \cdot C_{\text{L}}(m,n) \right] \qquad (42)$$

It is evident from the discussion related to Fig. 15(d) above that $S_{\text{mean}}$ at 600ºC and 700ºC is hypo-stoichiometric at low doses while it becomes hyper-stoichiometric at high dose. The irradiation dose at which this transition happens reduces as the temperature increases. Also, at low temperature of 400ºC, $S_{\text{mean}}$ is hypo-stoichiometric over the entire range of irradiation dose. Defect clustering leads to an imbalance of monomer concentration in the matrix rendering it off-stoichiometric. The off-stoichiometry of the matrix is expressed as $ThO_{2+x}$ where the stoichiometric parameter, $x$ is:

$$x = \frac{N_{\text{O}} - C_{V_{\text{O}}} + C_{\text{O}_i}}{N_{\text{Th}} - C_{V_{\text{Th}}} + C_{\text{Th}_i}} - 2 \qquad (43)$$

where $N_{\text{O}}$ and $N_{\text{Th}}$ is the density of O and Th atoms in the $ThO_2$ lattice and $C_{V_{\text{O}}}$, $C_{\text{O}_i}$, $C_{V_{\text{Th}}}$ and $C_{\text{Th}_i}$ are the concentration of respective monomers. From Fig. 16(a) and (b) we can clearly see that $V_{\text{Th}}$ is the dominant monomer in the matrix at irradiation doses greater than $10^{-5}$ dpa for all temperature. This is primarily due to the accumulation of Th vacancies on account of their sluggish kinetics which results in a hyper-stoichiometric matrix indicated by a positive value of $x$. The variation of matrix stoichiometry with dose at different temperature is shown in Fig. 16 (d). We see that the matrix remains hyper-stoichiometric over the entire range of irradiation dose at any specific temperature. The hyper-stoichiometry reduces with the increase in temperature because the concentration of $V_{\text{Th}}$ starts depleting even at lower dose due to accelerated clustering and monomer recombination reactions.

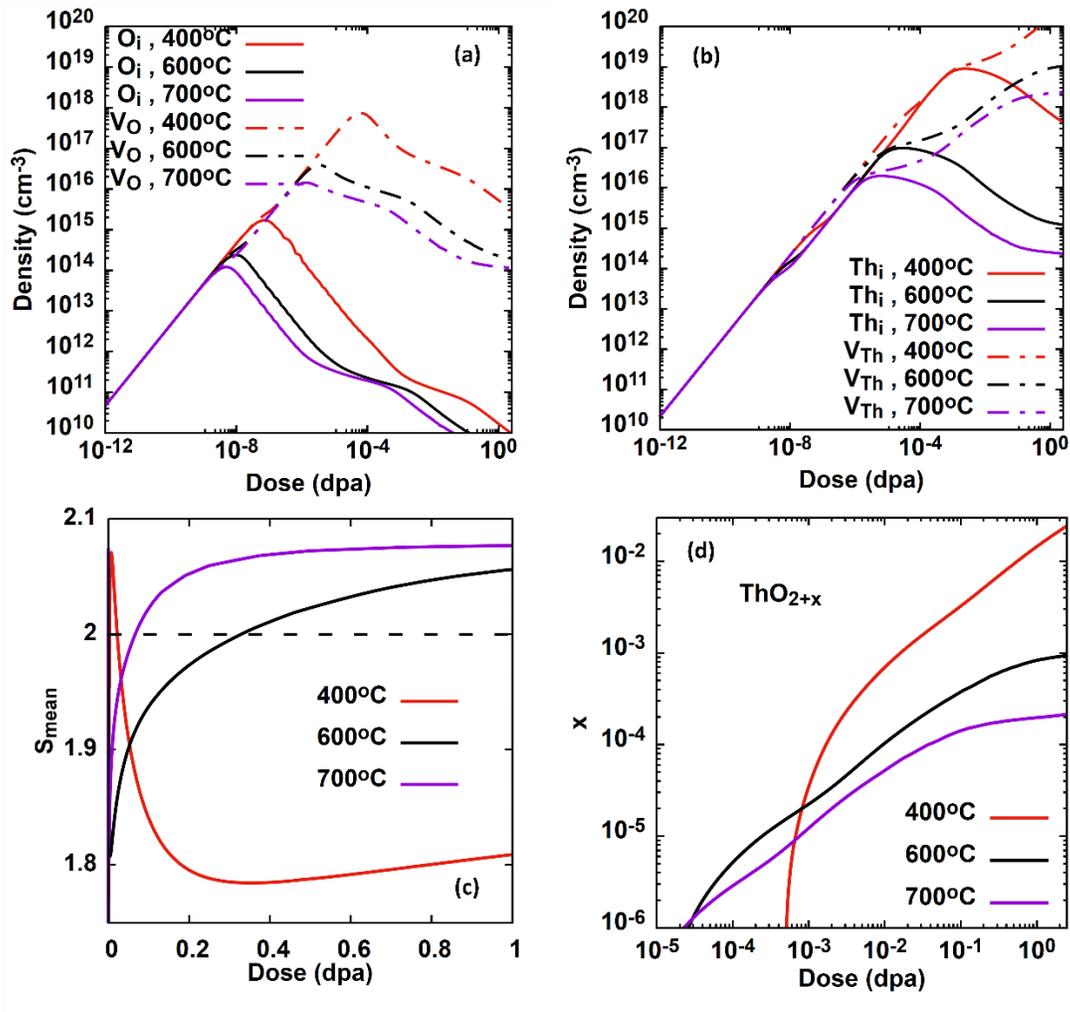

Fig. 16. Evolution of (a) O monomers and (b) Th monomers with dose at 400ºC, 600ºC and 700ºC. (c) Evolution of average loop stoichiometry, $S_{mean}$ with dose at different temperatures. (d) Evolution of the matrix stoichiometry denoted by $x$, with dose at different temperature.

## 4. Discussion

The cluster composition spaces (CCSs) depicted in Fig. 3 show the composition of all the crystallographically possible interstitial and vacancy clusters (loops, respectively, voids) in the ThO₂ fluorite lattice. To study the agglomeration of point defects into clusters having non-stoichiometric composition, we have investigated the evolution of loops and voids in their respective CCSs. The energetics and kinetics of the clustering phenomenon is primarily governed by the binding energy of monomers to clusters, i.e., $E_{Th_i}^b(n, m), E_{O_i}^b(n, m), E_{V_{Th}}^b(n, m)$ and $E_{V_O}^b(n, m)$ and the monomer diffusivities, i.e., $D_{Th_i}, D_{O_i}, D_{V_{Th}}$ and $D_{V_O}$ in the $ThO_2$ matrix, respectively. Mapping the binding energies on the respective CCS of loops and voids is insightful in understanding the preferential evolution of clusters over a selective region of the CCS. For example, in Fig. 10 (a) and (b) SIA loops $(n, m)$ in the CCS with negative $E_{O_i}^b(n, m)$ and $E_{Th_i}^b(n, m)$ are not likely to evolve due to high emission coefficients, $\alpha_{O_i}(n, m)$ and $\alpha_{Th_i}(n, m)$. Similarly, in Fig. 11 (a) and (b) voids $(n, m)$ with negative $E_{V_{Th}}^b(n, m)$ and $E_{V_O}^b(n, m)$ are unlikely to grow in the irradiated matrix. Thus, computing the cluster energetics across the entire range of stoichiometry is crucial in predicting the evolution of non-stoichiometric loops and voids during irradiation. This is a significant improvement over the simple assumption by Khalil et al. [20] that the binding energy of monomers is solely governed by the surface energy as in metals.

In addition to the energetics, exploring the configuration space of clusters is yet another essential contribution of this work. Clustering is a stochastic phenomenon specifically when the monomers begin agglomerating into small clusters of varying geometry, based on their arrangement in different configurations. Fig. 7(b) is a histogram plot which shows the clustering energy, $E_L^{Cluster}(n, m)$ of loops distributed across all possible configurations. Due to the computational expense, our investigation of the energies using MD was limited to small loops with 6 $Th_i$ and 25 $O_i$. The typical distribution of $E_L^{Cluster}(n, m)$ allows us to compute a mean value and fluctuations in $E_L^{Cluster}(n, m)$ about the mean, associated with the variable loop configuration at the same size and composition. The resulting fluctuation is a result of differences in lattice elastic distortion, surface (perimeter) energy and strong interaction with neighboring atoms due to varying loop geometry. It, however, seems to clearly increase with the loop size as is evident from Fig. 7. A detailed analysis of the uncertainties in $E_L^{Cluster}(n, m)$ is currently in progress to understand they can be incorporated in CD, thus enhancing the predictions in the probable trajectory of SIA clustering in the configuration space.

The monomer diffusivities in the irradiated matrix govern the clustering kinetics which determine the size and distribution of clusters in the final microstructure. Temperature accelerated MD simulations with the EAM potential by Cooper et al. [24] predicts the migration energy and pre-exponential diffusivity factors of $O_i$, $Th_i$ and $V_O$ as discussed comprehensively in Section 2.3.2. $O_i$ and $Th_i$ were observed to diffuse via interstitialcy mechanism in TAMD simulations. The predicted migration barriers of $O_i$ and $V_O$ differ from the CI-NEB calculations in DFT by ~0.54 eV and ~0.71 eV respectively. Migration energy of $Th_i$ obtained using accelerated MD is 3.172 eV which is smaller than cation migration energies in $UO_2$ and $CeO_2$ calculated using DFT. Migration of $V_{Th}$ is predicted to be significantly sluggish and hence not captured by TAMD even at very high temperature of 2800 K. However, we have used the experimental data on self-diffusivity of Th by Matzke [38], equilibrium concentration of $Th_i$ and $V_{Th}$ from thermodynamic calculations by Maniesha et al. [39] and $E_{Th_i}^m$ and $D_{Th_i}^o$ obtained from accelerated MD, to have a reasonable estimate of the migration energy, $E_{V_{Th}}^m$ and the pre-exponential diffusivity factor, $D_{V_{Th}}^o$ from Eq. (38). The estimated migration energy, i.e., 4.837 eV is adequately close to the DFT result, i.e., 4.47 eV [31] but most importantly we have an estimate of $D_{V_{Th}}^o$ that can be improved in future studies.

The cluster dynamics model with its reference set of parameters yield the density of monomers and clusters with different irradiation dose at 600ºC. However, CD predicted the density of large loops to deviate significantly from the TEM observations. On a careful analysis, this discrepancy was reasonably attributed to the uncertainty associated with the estimated parameters. Accuracy of the computed point defect diffusivities as discussed above rely on the limitations of the EAM potential used for the TAMD simulations. Whereas, for the binding energy of monomers to clusters, in addition to the limitations of EAM potential, probability distribution across all possible cluster configurations define the uncertainty associated with it. The sensitivity of our CD model to the parameters mentioned above justifies optimizing them to better fit the TEM observations. Range of uncertainty in binding energies calculated from the uncertainty of $E_{\text{L}}^{\text{Cluster}}(n, m)$ of loops associated with the choice of loop geometry, defines their window of optimization. Similarly, the defect migration energies were incrementally varied on either side of their reference values until the CD predicted SDF of large loops fitted the TEM observations closely. The final set of optimized parameters are discussed comprehensively in Section 3. It is interesting to note that optimized values of $E_{\text{O}_i}^{\text{m}}$ and $E_{\text{V}_\text{O}}^{\text{m}}$, i.e., 1 eV and 1.5 eV are very close to the results from CI-NEB calculations in DFT, i.e., 1.04 eV [29] and 1.27 eV [31] respectively. Again, this raises an issue as to the suitability of the EAM potential by Cooper et al. in predicting the mobility of monomers in $ThO_2$.

The cluster dynamics model with its fitted set of parameters calculate the density evolution of loops and monomers in the irradiated $ThO_2$ matrix. As discussed in Section 3, results from the CD model, i.e., the SDF of large loops, average loop radius and total loop density, at different irradiation doses and 600ºC closely corresponds to the TEM observations. The CD calculated total loop density is of the same order of magnitude but almost three times larger than the observations. This discrepancy is reasonably attributed to firstly, the error associated with the choice of parameters, but most importantly, the limited resolution of TEM in accurately measuring the density of small loops. Thus, the objective of the model is to accurately predict the density of small loops and monomers that are unresolved in TEM. In addition to the available experimental data, large loop density observed in samples irradiated under a range of conditions, i.e., temperature and dose rate can be used to optimize the parameters even better. Characteristic loop density distribution obtained from annealing irradiated microstructures, can also be used for further parameter optimization. Thus, future work on the model is primarily focused on obtaining the best fit parameters that would enhance the fidelity of CD predictions.

## 5. Concluding Remarks

In this work we have used an atomistically informed cluster dynamics model to predict the nucleation and growth kinetics of point defect clusters, i.e., SIA loops and voids, in irradiated $ThO_2$. The composition spaces of clusters were determined from the geometry of a $ThO_2$ fluorite lattice. Using atomistic simulations, we have obtained reasonable estimates of the inputs to our CD model, i.e., the binding energy of point defects to corresponding clusters and the diffusivity of individual monomers. The energetics of all possible SIA loop configurations expected during cluster nucleation have also been explored. An optimized set of input parameters were obtained by varying them within their windows of uncertainty and checking for the best fit of CD predictions with TEM observations.

 The nucleation and growth of SIA loops having near-stoichiometric compositions is favoured due to the cluster energetics. At high irradiation temperatures of 600ºC and 700ºC, the loops are predominantly hypo-stoichiometric at low doses. Whereas, at high doses the density of hyper-stoichiometric loops takes over. At low temperature, i.e., 400ºC the loops remain predominantly hypo-stoichiometric at any dose. At 600ºC, the size distribution of loops at different irradiation doses closely corresponds to the TEM observation reported by Dennett, Deskins and Khafizov et al. [40]. The evolution of average loop radius with dose also falls within the error of experimental observations. The total density of loops predicted by CD seems to be slightly higher because small loops are difficult to resolve accurately in TEM. Thus, the notable utility of this model lies in predicting the density of monomers and small loops having radii less than 1 nm with a certain degree of fidelity. As mentioned before, this information is crucial in modelling the phenomenon of phonon scattering by microstructural defects to accurately predict the thermal properties of irradiated $ThO_2$.

## Conflicts of interest

There are no conflicts to declare.

## Acknowledgements

This work was supported by Center for Thermal Energy Transport under Irradiation, an Energy Frontier Research Center funded by the U.S. Department of Energy, Office of Science, Office of Basic Energy Sciences. Tomohisa Kumagai was supported by the Central Research Institute of Electric Power Industry (CRIEPI), Yokosuka, Kanagawa, 2400196, Japan. We would like to acknowledge the valuable inputs from Joseph Anderson on the configuration space of SIA loops and Kyle Starkey while building the cluster dynamics code.

## References

1    M. Mann, D. Thompson, K. Serivalsatit, T. M. Tritt, J. Ballato and J. Kolis, *Crystal Growth and Design*, 2010, 10, 2146–2151.

2    K. Bakker, E. H. P. Cordfunke, R. J. M. Konings and R. P. C. Schram, *Journal of Nuclear Materials*, 1997, 250, 1-12.

3    C. A. Dennett, Z. Hua, A. Khanolkar, T. Yao, P. K. Morgan, T. A. Prusnick, N. Poudel, A.

French, K. Gofryk, L. He, L. Shao, M. Khafizov, D. B. Turner, J. M. Mann and D. H. Hurley, *APL Mater.*, 2020, 8, 111103.

4    M. Khafizov, V. Chauhan, Y. Wang, F. Riyad, N. Hang and D. H. Hurley, *J. Mater. Res.*, 2017, 32, 204–216.

5    M. Khafizov, J. Pakarinen, L. He and D. H. Hurley, *J. Am. Ceram. Soc.*, 2019, 102, 7533–7542.

6    L. K. Mansur, *J. Nucl. Mater.*, 1994, 216, 97-123.

7    A. H. Duparc, C. Moingeon, N. Smetniansky-De-Grande and A. Barbu, *J. Nucl. Mater.*, 2002, 302, 143-155.

8    K. Radhakrishnan and A. C. Hindmarsh, *NASA Reference Publication 1327, Lawrence Livermore Nationai Laboratory Report UCRL-ID-113855,* 1993.

9    S. D. Cohen and A. C. Hindmarsh, *Comput. Phys.*, 1996, 10, 138–143.

10   A. Gokhman and F. Bergner, *Radiat. Eff. Defects Solids*, 2010, 165, 216–226.

11   C. Pokor, Y. Brechet, P. Dubuisson, J. P. Massoud and A. Barbu, *J. Nucl. Mater.*, 2004, 326, 19–29.

12   D. Brimbal, L. Fournier and A. Barbu, *J. Nucl. Mater.*, 2016, 468, 124–139.

13   F. Christien and A. Barbu, *J. Nucl. Mater.*, 2005, 346, 272–281.

14   E. Clouet, A. Barbu, L. Laé and G. Martin, *Acta Mater.*, 2005, 53, 2313–2325.

15   X. M. Bai, H. Ke, Y. Zhang and B. W. Spencer, *J. Nucl. Mater.*, 2017, 495, 442–454.

16   E. Clouet, M. Nastar, A. Barbu, C. Sigli and G. Martin, arXiv, 2005, preprint, <u>arXiv:cond-</u>

mat/0507259v2, https://arxiv.org/abs/cond-mat/0507259

17  A. R. Hassan, A. El-Azab, C. Yablinsky and T. Allen, *J. Solid State Chem.*, 2013, 204, 136–145.

18  R. Skorek, S. Maillard, A. Michel, G. Carlot, E. Gilabert and T. Jourdan, *Defect and Diffusion Forum*, 2012, 323–325, 209–214.

19  J. Jonnet, P. Van Uffelen, T. Wiss, D. Staicu, B. Rémy and J. Rest, *Nucl. Instruments Methods Phys. Res. Sect. B Beam Interact. with Mater. Atoms*, 2008, 266, 3008–3012.

20  S. Khalil, T. Allen and A. El-Azab, *Chem. Phys.*, 2017, 487, 1–10.

21  N. Soneda and T. Diaz De La Rubia, *Philos. Mag. A Phys. Condens. Matter, Struct. Defects Mech. Prop.*, 1998, 78, 995–1019.

22  K. Bawane, X. Liu, T. Yao, M. Khafizov, A. French, J. M. Mann, L. Shao, J. Gan, D. H. Hurley and L. He, *J. Nucl. Mater.*, 2021, 552, 152998.

23  A. P. Thompson, H. M. Aktulga, R. Berger, D. S. Bolintineanu, W. M. Brown, P. S. Crozier, P. J. in 't Veld, A. Kohlmeyer, S. G. Moore, T. D. Nguyen, R. Shan, M. J. Stevens, J. Tranchida, C. Trott and S. J. Plimpton, *Comput. Phys. Commun.*, 2022, 271, 108171.

24  M. W. D. Cooper, M. J. D. Rushton and R. W. Grimes, *J. Phys. Condens. Matter*, 2014, 26, 105401.

25  A. F. Voter and M. R. Sorensen, *Mater. Res. Soc. Symp. - Proc.*, 1999, 538, 427–439.

26  A. F. Voter, *Phys. Rev. Lett.*, 1997, 78, 3908

27  R. A. Miron and K. A. Fichthorn, *J. Chem. Phys.*, 2003, 119, 6210–6216.

28    A. Stukowski, *Model. Simul. Mater. Sci. Eng.,* 2009, 18, 015012.

29    H. Y. Xiao, Y. Zhang and W. J. Weber, *Acta Mater.*, 2013, 61, 7639–7645.

30    D. S. Aidhy, *Phys. Chem. Chem. Phys.*, 2016, 18, 15019–15024.

31    Y. Yun, P. M. Oppeneer, H. Kim and K. Park, *Acta Mater.*, 2009, 57, 1655–1659.

32    H. Y. Xiao and W. J. Weber, *J. Phys. Chem. B*, 2011, 115, 6524–6533.

33    B. Liu, D. S. Aidhy, Y. Zhang and W. J. Weber, *Phys. Chem. Chem. Phys.*, 2014, 16, 25461–25467.

34    B. Dorado, D. A. Andersson, C. R. Stanek, M. Bertolus, B. P. Uberuaga, G. Martin, M. Freyss and P. Garcia, *Phys. Rev. B - Condens. Matter Mater. Phys.*, 2012, 86, 1–10.

35    S. Beschnitt, T. Zacherle and R. A. De Souza, *J. Phys. Chem. C*, 2015, 119, 27307–27315.

36    H. Y. Xiao, Y. Zhang and W. J. Weber, *J. Nucl. Mater.*, 2011, 414, 464–470.

37    A. D. King, *J. Nucl. Mater.*, 1971, 38, 347–349.

38    H. Matzke, *Philos. Mag. A Phys. Condens. Matter, Struct. Defects Mech. Prop.*, 1991, 64, 1181–1200.

39    M. Singh, T. Kumagai, J. Wharry and A. El-Azab, *J. Nucl. Mater.,* 2022, 567, 153804.

40    C. A. Dennett, W. R. Deskins, M. Khafizov, Z. Hua, A. Khanolkar, K. Bawane, L. Fu, J. M. Mann, C. A. Marianetti, L. He, D. H. Hurley and A. El-Azab, *Acta Mater.*, 2021, 213, 116934.

# Atomistically-informed modeling of point defect clustering and evolution in irradiated ThO2

Sanjoy Kumar Mazumder [a, *], Maniesha Kaur Salaken Singh [b], Tomohisa Kumagai [c], Anter El-Azab [a]

[a] School of Materials Engineering, Purdue University
West Lafayette, Indiana 47096 (US)
E-mail: mazumder@purdue.edu

[b] School of Nuclear Engineering, Purdue University
West Lafayette, Indiana 47096 (US)

[c] Central Research Institute of Electric Power Industry (CRIEPI)
Yokosuka, Kanagawa 2400196 (Japan)

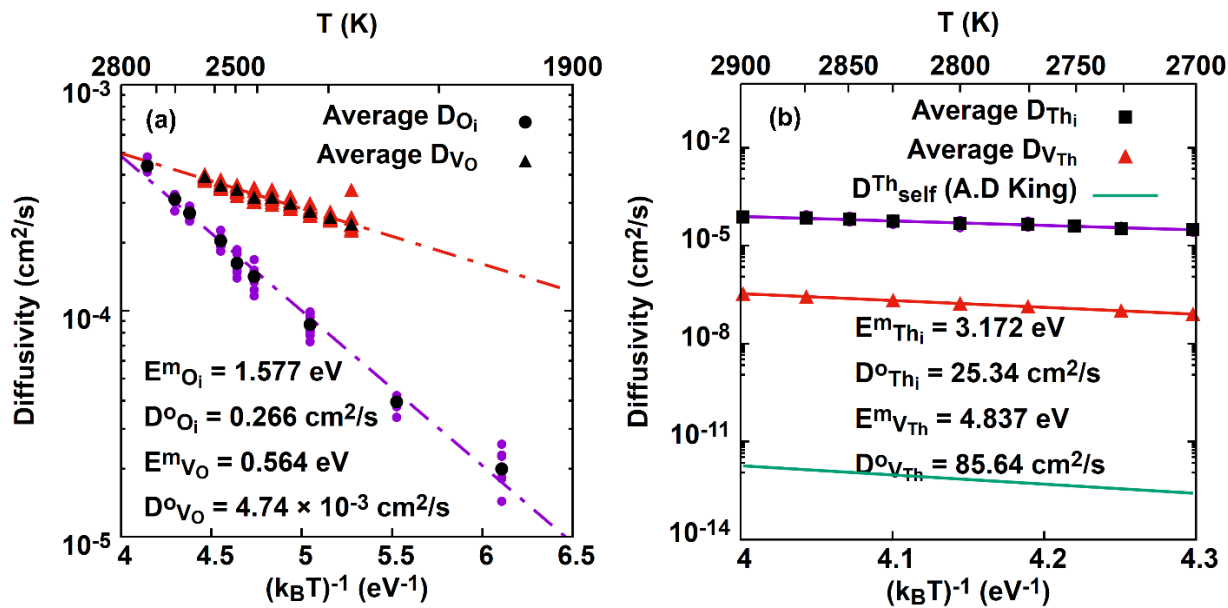

*Figure S1.* Plots showing the diffusivity of monomers obtained from TAMD (a) $D_{O_i}$ and $D_{V_O}$ (b) $D_{Th_i}$, $D_{Th}^{self}$ [1]) and calculated $D_{V_{Th}}$.

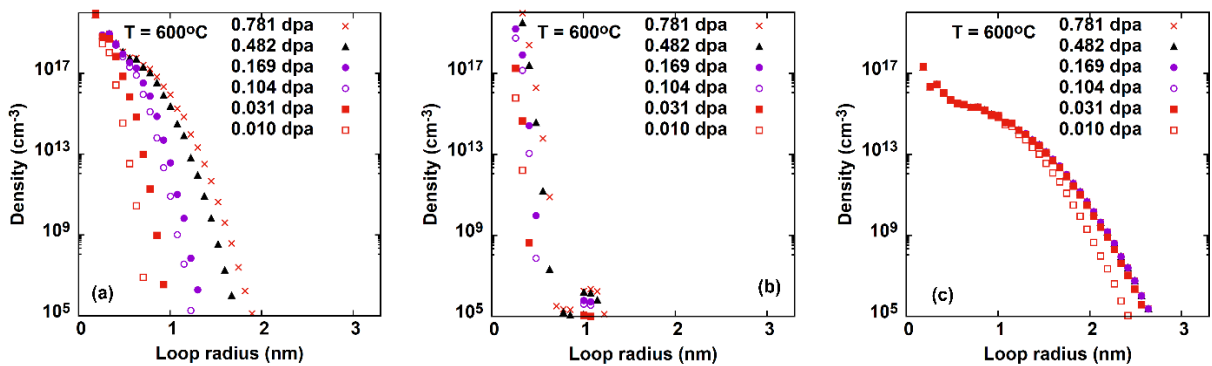

*Figure S2.* CD results showing the SDF of loops obtained with (a) the TAMD diffusivity, i.e., $E_{O_i}^m = 1.577$ eV, $E_{V_O}^m = 0.564$ eV, $E_{Th_i}^m = 3.172$ eV and $E_{V_{Th}}^m = 4.837$ eV (b) altered $E_{Th_i}^m = 3.8$ eV (c) altered $E_{Th_i}^m = 2.3$ eV.

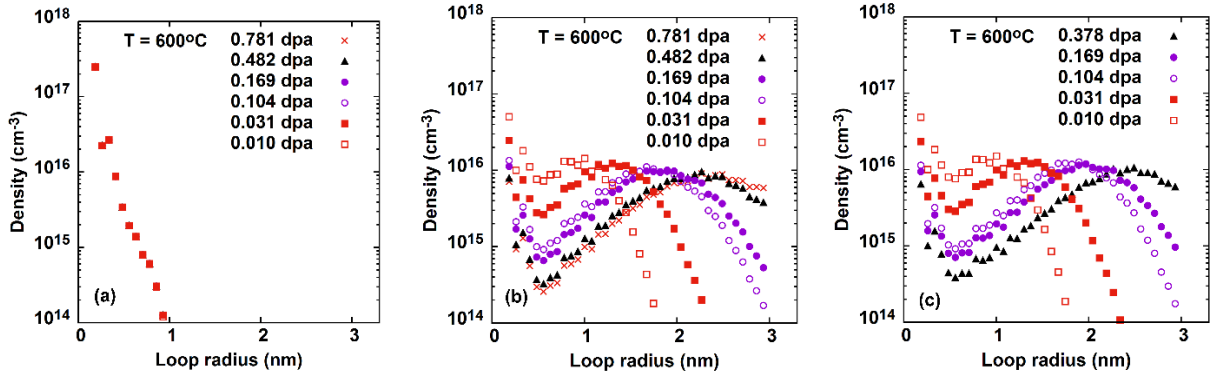

*Figure S3.* CD results showing the SDF of loops obtained with (a) $E_{O_i}^m = 1.577$ eV, $E_{V_O}^m = 0.3$ eV, $E_{Th_i}^m = 2.3$ eV and $E_{V_{Th}}^m = 4.837$ eV (b) $E_{V_O}^m = 1$ eV (c) $E_{V_O}^m = 1.5$ eV.

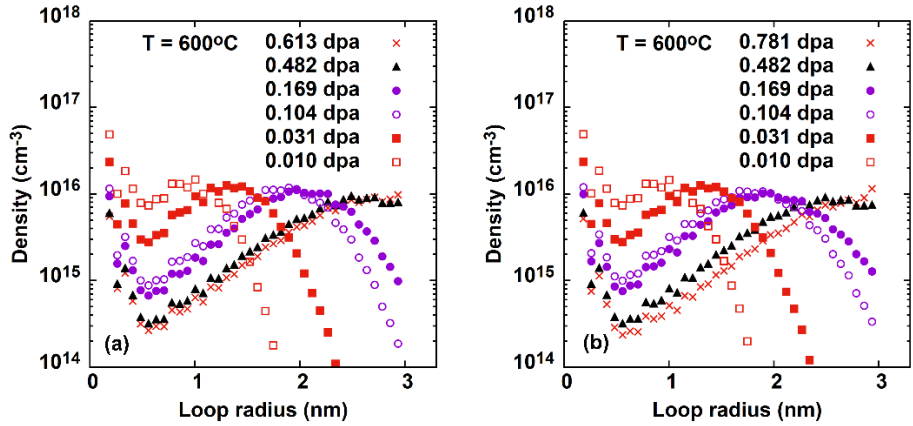

*Figure S4.* CD results showing the SDF of loops obtained with (a) $E_{O_i}^m = 0.71$ eV, $E_{V_O}^m = 1.5$ eV, $E_{Th_i}^m = 2.3$ eV and $E_{V_{Th}}^m = 4.837$ eV (b) $E_{O_i}^m = 1$ eV.

Reference

1) A. D. King, *J. Nucl. Mater.*, 1971, 38, 347–349.